\documentstyle[prb,aps]{revtex}
\begin{document}

\title{Bosonization Rules for Electron-Hole Systems - II}
\author{Girish S. Setlur}
\address{Center for Laser and Photonics Research, Oklahoma State 
 University, Stillwater, OK}
\maketitle

\begin{abstract}
 Here we write down and prove closed commutation rules for Fermi bilinears
 in a two-component Fermi system in terms of the relevant 
 sea-bosons. We show how the commutation rules come out correctly
 within the RPA-approximation as do the zero temperature correlation functions.
 We write down hamiltonains of nonrelativistic interacting electron-hole
 systems and point out several attractive features such as the natural 
 roles excitons play in this language. We then use this language to derive
 a set of equations analogous to the Semiconductor Bloch Equations
 in the presence of phonons and with external fields such as those present
 in pump-probe experiments. We solve these equations using parameters
 of interesting and topical materials
 such as GaN and compare with some recent experimental data.
\end{abstract}

\section{Bilinear Sea-boson Correspondence}

 Here we write down the formal correspondence between Fermi bilinears and
 sea-bosons suitably generalised to two-component systems.
 The correspondence presented here is necessarily approximate but one that 
 conforms to the spirit of the random-phase approximation, which in the 
 two-component system corresponds to the exciton approximation. However,
 just as in the one-component case, new physics could be extracted by suitably
 generalising the notion of the random-phase approximation, we find that here
 too we may extract new physics by generalising the notion of the
 exciton approximation so that one goes beyond the dilute limit and includes
 experimentally relevent situations such as those in which a
 large number of real carriers are created by external fields. 
 The discussion presented here follows closely the discussion
 in the response to the comment on our previous work\cite{Setlur}. 
 Let us write the Fermi bilinear sea-boson correspondence with spin.
 ($ {\bf{q}} \neq 0 $)
\[
c^{\dagger}_{ {\bf{k}} + {\bf{q}}/2, \sigma }
c_{ {\bf{k}} - {\bf{q}}/2, \sigma^{'} }
 = \Lambda_{ {\bf{k}}, \sigma }({\bf{q}}, \sigma^{'})
a_{ {\bf{k}}, \sigma }(-{\bf{q}}, \sigma^{'})
+ a^{\dagger}_{ {\bf{k}}, \sigma^{'} }({\bf{q}}, \sigma)
\Lambda_{ {\bf{k}}, \sigma^{'} }(-{\bf{q}}, \sigma)
\]
\[
+ \sum_{ {\bf{q}}_{1}\sigma_{1} }
a^{\dagger}_{ {\bf{k+q/2-q_{1}/2}}\sigma_{1} }({\bf{q_{1}}}\sigma)
a_{ {\bf{k-q_{1}/2}}\sigma_{1} }({\bf{q_{1}-q}}\sigma^{'})
\]
\begin{equation}
-  \sum_{ {\bf{q}}_{1}\sigma_{1} }
 a^{\dagger}_{ {\bf{k-q/2+q_{1}/2}}\sigma^{'} }({\bf{q_{1}}}\sigma_{1})
a_{ {\bf{k+q_{1}/2}}\sigma }({\bf{q_{1}-q}}\sigma_{1})
\end{equation}
Here,
\begin{equation}
 \Lambda_{ {\bf{k}}, \sigma }({\bf{q}}, \sigma^{'}) =
\sqrt{
 {\bar{n}}_{ {\bf{k}} + {\bf{q}}/2, \sigma }
(1 - {\bar{n}}_{ {\bf{k}} - {\bf{q}}/2, \sigma^{'} })}
\end{equation}
 Let us make the following identifications,
\begin{equation}
c_{ {\bf{k}} \uparrow } = c_{ {\bf{k}} }
\end{equation}
\begin{equation}
c_{ {\bf{k}} \downarrow } = d^{\dagger}_{ -{\bf{k}} }
\end{equation}
 Taking a cue from the one-component case let us now argue that( for both
 $ {\bf{q}} = 0 $ and $ {\bf{q}} \neq 0 $)
\[
d_{ -{\bf{k}}-{\bf{q}}/2 }c_{ {\bf{k}} - {\bf{q}}/2 }
 \approx \sqrt{ (1 - {\bar{n}}_{e}({\bf{k}} - {\bf{q}}/2))
( 1 - {\bar{n}}_{h}(-{\bf{k}} - {\bf{q}}/2) ) }
a_{ {\bf{k}}\downarrow }(-{\bf{q}}\uparrow)
\]
\begin{equation}
 + \sqrt{ {\bar{n}}_{e}({\bf{k}} - {\bf{q}}/2)
{\bar{n}}_{h}(-{\bf{k}} - {\bf{q}}/2) }
a^{\dagger}_{ {\bf{k}}\uparrow }({\bf{q}}\downarrow)
\label{OFFDIAG}
\end{equation}
\begin{equation}
\Lambda_{1}({\bf{k}},{\bf{q}})
 = \sqrt{ (1 - {\bar{n}}_{e}({\bf{k}} - {\bf{q}}/2))
( 1 - {\bar{n}}_{h}(-{\bf{k}} -{\bf{q}}/2) ) }
\end{equation}
\begin{equation}
\Lambda_{2}({\bf{k}},{\bf{q}}) = 
 \sqrt{ {\bar{n}}_{e}({\bf{k}} - {\bf{q}}/2)
{\bar{n}}_{h}(-{\bf{k}} -{\bf{q}}/2) }
\end{equation}
 The above form in Eq.(~\ref{OFFDIAG}) automatically satisfies,
\begin{equation}
[d_{ -{\bf{k}}-{\bf{q}}/2 }c_{ {\bf{k}} - {\bf{q}}/2 },
d_{ -{\bf{k}}^{'}-{\bf{q}}^{'}/2 }c_{ {\bf{k}}^{'} - {\bf{q}}^{'}/2 }] = 0
\end{equation}
 Next we would like the following to happen (for both $ {\bf{q}} = 0 $ 
 and $ {\bf{q}} \neq 0 $), as it does in the Fermi language,
\[
[d_{ -{\bf{k}}-{\bf{q}}/2 }c_{ {\bf{k}} - {\bf{q}}/2 },
c^{\dagger}_{ {\bf{k}}^{'} - {\bf{q}}^{'}/2 }
d^{\dagger}_{ -{\bf{k}}^{'}-{\bf{q}}^{'}/2 }] =
\]
\[
\delta_{ {\bf{k}}, {\bf{k}}^{'} }\delta_{ {\bf{q}}, {\bf{q}}^{'} }
- d^{\dagger}_{ -{\bf{k}}^{'}-{\bf{q}}^{'}/2 }d_{ -{\bf{k}}-{\bf{q}}/2 }
\delta_{ {\bf{k}} - {\bf{q}}/2, {\bf{k}}^{'} - {\bf{q}}^{'}/2 }
 - c^{\dagger}_{ {\bf{k}}^{'} - {\bf{q}}^{'}/2 }c_{ {\bf{k}} - {\bf{q}}/2 }
\delta_{ {\bf{k}} + {\bf{q}}/2, {\bf{k}}^{'} + {\bf{q}}^{'}/2 }
\]
\begin{equation}
\approx \delta_{ {\bf{k}}, {\bf{k}}^{'} }\delta_{ {\bf{q}}, {\bf{q}}^{'} }
( 1 - \langle 
 d^{\dagger}_{ -{\bf{k}}-{\bf{q}}/2 }d_{ -{\bf{k}}-{\bf{q}}/2 } \rangle 
 - \langle c^{\dagger}_{ {\bf{k}} - {\bf{q}}/2 }
c_{ {\bf{k}} - {\bf{q}}/2 } \rangle )
\end{equation}
\begin{equation}
(1 - {\bar{n}}_{e}({\bf{k}}-{\bf{q}}/2))
(1 - {\bar{n}}_{h}(-{\bf{k}}-{\bf{q}}/2))
 - {\bar{n}}_{e}({\bf{k}}-{\bf{q}}/2){\bar{n}}_{h}(-{\bf{k}}-{\bf{q}}/2)
 =  1 - \langle d^{\dagger}_{ -{\bf{k}}-{\bf{q}}/2 }
d_{ -{\bf{k}}-{\bf{q}}/2 } \rangle
 - \langle c^{\dagger}_{ {\bf{k}} - {\bf{q}}/2 }
c_{ {\bf{k}} - {\bf{q}}/2 } \rangle  
\end{equation}
 The expectation value is with respect to the full interacting ground state 
 including(especially including) the external fields that allow for
 significant real populations to be generated.
 Let us now argue(again inspired by the one-component system),
\begin{equation}
d^{\dagger}_{ -{\bf{k}} }d_{ -{\bf{k}} } = 
n^{(0)}_{h}({\bf{k}}) 
 - \sum_{ {\bf{q}}_{1} \sigma_{1} }
a^{\dagger}_{ {\bf{k}} - {\bf{q}}_{1}/2 \sigma_{1} }
({\bf{q}}_{1}\downarrow)
a_{ {\bf{k}} - {\bf{q}}_{1}/2\sigma_{1} }({\bf{q}}_{1}\downarrow)
 + \sum_{ {\bf{q}}_{1} \sigma_{1} }
a^{\dagger}_{ {\bf{k}} + {\bf{q}}_{1}/2\downarrow }
({\bf{q}}_{1}\sigma_{1})
a_{ {\bf{k}} + {\bf{q}}_{1}/2 \downarrow }({\bf{q}}_{1}\sigma_{1})
\end{equation}
and,
\begin{equation}
c^{\dagger}_{ {\bf{k}} }c_{ {\bf{k}} } =
n^{(0)}_{e}({\bf{k}}) + 
 \sum_{ {\bf{q}}_{1} \sigma_{1} }
a^{\dagger}_{ {\bf{k}} - {\bf{q}}_{1}/2 \sigma_{1} }
({\bf{q}}_{1}\uparrow)
a_{ {\bf{k}} - {\bf{q}}_{1}/2\sigma_{1} }({\bf{q}}_{1}\uparrow)
 - \sum_{ {\bf{q}}_{1} \sigma_{1} }
a^{\dagger}_{ {\bf{k}} + {\bf{q}}_{1}/2\uparrow }
({\bf{q}}_{1}\sigma_{1})
a_{ {\bf{k}} + {\bf{q}}_{1}/2 \uparrow }({\bf{q}}_{1}\sigma_{1})
\end{equation}
 Here $ n^{(0)}_{h}({\bf{k}}) $  and $  n^{(0)}_{e}({\bf{k}}) $
 account for possible doping in the system. In other words, the presence of
 excess charge. 
 These two put together obey the attractive identity(charge conservation),
\begin{equation}
\sum_{ {\bf{k}} } c^{\dagger}_{ {\bf{k}} }c_{ {\bf{k}} }
 - \sum_{ {\bf{k}} } d^{\dagger}_{ -{\bf{k}} }d_{ -{\bf{k}} } = Q
 = \sum_{ {\bf{k}} }[n^{(0)}_{e}({\bf{k}})-n^{(0)}_{h}({\bf{k}})]
\end{equation}
 Now we move on to the off-diagonal(in the indices)
 parts($ {\bf{q}} \neq 0 $),
\begin{equation}
d^{\dagger}_{ -{\bf{k}} + {\bf{q}}/2 }d_{ -{\bf{k}} - {\bf{q}}/2 }
 = -\Lambda_{h}({\bf{k}},{\bf{q}})
a_{ {\bf{k}} \downarrow }(-{\bf{q}}\downarrow)
 - \Lambda_{h}({\bf{k}},-{\bf{q}})
a^{\dagger}_{ {\bf{k}} \downarrow }({\bf{q}}\downarrow)
\end{equation}
\begin{equation}
c^{\dagger}_{ {\bf{k}} + {\bf{q}}/2 }c_{ {\bf{k}} - {\bf{q}}/2 }
 = \Lambda_{e}({\bf{k}},{\bf{q}})a_{ {\bf{k}}\uparrow }
(-{\bf{q}}\uparrow)
 + \Lambda_{e}({\bf{k}},-{\bf{q}})a^{\dagger}_{ {\bf{k}}\uparrow }
({\bf{q}}\uparrow)
\end{equation}
 Assuming that the ground states(of the non-interacting system)
 are annhilated by the sea-bosons
 $ a_{ {\bf{k}} \downarrow }({\bf{q}}\downarrow) $
 and $ a_{ {\bf{k}} \uparrow }({\bf{q}}\uparrow) $
\begin{equation}
a_{ {\bf{k}} \downarrow }({\bf{q}}\downarrow) |Free \rangle = 0
,\mbox{                     }
a_{ {\bf{k}} \uparrow }({\bf{q}}\uparrow) |Free \rangle
 = 0
\end{equation}
This means,
\begin{equation}
\langle c^{\dagger}_{ {\bf{k}} + {\bf{q}}/2 }c_{ {\bf{k}} - {\bf{q}}/2 }
c^{\dagger}_{ {\bf{k}} - {\bf{q}}/2 }c_{ {\bf{k}} + {\bf{q}}/2 } \rangle
 = {\bar{n}}_{e}({\bf{k}} + {\bf{q}}/2)
(1 - {\bar{n}}_{e}({\bf{k}} - {\bf{q}}/2)  ) 
 = \Lambda^{2}_{e}({\bf{k}},{\bf{q}})
\end{equation}
and similarly for the holes.
\begin{equation}
\Lambda_{e}({\bf{k}},{\bf{q}}) = 
\sqrt{ {\bar{n}}_{e}({\bf{k}} + {\bf{q}}/2)
(1 - {\bar{n}}_{e}({\bf{k}} - {\bf{q}}/2)  ) }
\end{equation}
\begin{equation}
\Lambda_{h}({\bf{k}},{\bf{q}}) = 
\sqrt{ {\bar{n}}_{h}(-{\bf{k}} + {\bf{q}}/2)
(1 - {\bar{n}}_{h}(-{\bf{k}} - {\bf{q}}/2)  ) }
\end{equation}
 Then we make a leap of faith and suggest that the same should hold even
 when there are inteactions present and even when external fields are
 present. That is, the
 $ {\bar{n}}_{h}({\bf{k}}) = \langle d^{\dagger}_{ -{\bf{k}} }d_{ -{\bf{k}} } \rangle $
 now represents the expectation value with respect to the full interacting ground state. Lastly
 we would like to point out the internal self-consistency of this approach
 by computing the commutator bettween the diagonal and the off-diagonal 
 bilinears. We find much to our relief that no matter what the choices
 for the cofficients
 $ \Lambda_{1} $, $ \Lambda_{2} $, $ \Lambda_{e} $ and $  \Lambda_{h} $ are,
 we recover the following exact identities.
\begin{equation}
[d_{ -{\bf{k}} - {\bf{q}}/2 }c_{ {\bf{k}} - {\bf{q}}/2 },
 c^{\dagger}_{ {\bf{p}} }c_{ {\bf{p}} }]
 = d_{ -{\bf{k}} - {\bf{q}}/2 }c_{ {\bf{k}} - {\bf{q}}/2 }
\delta_{ {\bf{p}}, {\bf{k}} - {\bf{q}}/2 }
\end{equation}
\begin{equation}
[d_{ -{\bf{k}} - {\bf{q}}/2 }c_{ {\bf{k}} - {\bf{q}}/2 },
 d^{\dagger}_{ -{\bf{p}} }d_{ -{\bf{p}} }]
 = d_{ -{\bf{k}} - {\bf{q}}/2 }c_{ {\bf{k}} - {\bf{q}}/2 }
\delta_{ {\bf{p}}, {\bf{k}} + {\bf{q}}/2 }
\end{equation}
\begin{equation}
[c^{\dagger}_{ {\bf{k}} + {\bf{q}}/2 }c_{ {\bf{k}} - {\bf{q}}/2 },
c^{\dagger}_{ {\bf{p}} }c_{ {\bf{p}} }]
 = c^{\dagger}_{ {\bf{k}} + {\bf{q}}/2 }c_{ {\bf{k}} - {\bf{q}}/2 }
(\delta_{ {\bf{p}}, {\bf{k}} - {\bf{q}}/2 }
 - \delta_{ {\bf{p}}, {\bf{k}} + {\bf{q}}/2 })
\end{equation}
\begin{equation}
[d^{\dagger}_{ -{\bf{k}} + {\bf{q}}/2 }d_{ -{\bf{k}} - {\bf{q}}/2 },
d^{\dagger}_{ -{\bf{p}} }d_{ -{\bf{p}} }]
 = d^{\dagger}_{ -{\bf{k}} + {\bf{q}}/2 }d_{ -{\bf{k}} - {\bf{q}}/2 }
(\delta_{ {\bf{p}}, {\bf{k}} + {\bf{q}}/2 }
 - \delta_{ {\bf{p}}, {\bf{k}} - {\bf{q}}/2 })
\end{equation}
 Thus we have written down a potentially useful set of identities.
 Let us write down the hamiltonian of free electrons and holes.
 In the Fermi language it is,
\begin{equation}
 H_{free} = \sum_{ {\bf{k}} }\epsilon^{e}({\bf{k}})
 c^{\dagger}_{ {\bf{k}} }c_{ {\bf{k}} }
 + \sum_{ {\bf{k}} }\epsilon^{h}({\bf{k}})
 d^{\dagger}_{ -{\bf{k}} }d_{ -{\bf{k}} }
 + \sum_{ {\bf{q}} }\Omega_{LO} b^{\dagger}_{ {\bf{q}} }b_{ {\bf{q}} }
\end{equation}
The kinetic energy of the LO-phonon modes is also included.
In the sea-boson language it may be expressed as follows :
\[
H_{free} = -\sum_{ {\bf{k}}{\bf{q}} \neq {\bf{0}} }
(\epsilon^{e}({\bf{k}}-{\bf{q}}/2)
 + \epsilon^{h}({\bf{k}}+{\bf{q}}/2))
a^{\dagger}_{ {\bf{k}}\uparrow }({\bf{q}}\downarrow)
a_{ {\bf{k}}\uparrow }({\bf{q}}\downarrow)
\]
\[
 -\sum_{ {\bf{k}} }
(\frac{ k^{2} }{2\mu} + E_{g})
a^{\dagger}_{ {\bf{k}}\uparrow }({\bf{0}}\downarrow)
a_{ {\bf{k}}\uparrow }({\bf{0}}\downarrow)
+ \sum_{ {\bf{k}}{\bf{q}} \neq {\bf{0}} }(\epsilon^{h}({\bf{k}}-{\bf{q}}/2)
 - \epsilon^{h}({\bf{k}}+{\bf{q}}/2))
a^{\dagger}_{ {\bf{k}}\downarrow }({\bf{q}}\downarrow)
a_{ {\bf{k}}\downarrow }({\bf{q}}\downarrow)
\]
\[
+\sum_{ {\bf{k}}{\bf{q}} \neq {\bf{0}} }(\epsilon^{h}({\bf{k}}-{\bf{q}}/2)
 + \epsilon^{e}({\bf{k}}+{\bf{q}}/2))
a^{\dagger}_{ {\bf{k}}\downarrow}({\bf{q}}\uparrow)
a_{ {\bf{k}}\downarrow}({\bf{q}}\uparrow)
+\sum_{ {\bf{k}} }(\frac{ k^{2} }{2\mu} + E_{g})
a^{\dagger}_{ {\bf{k}}\downarrow}({\bf{0}}\uparrow)
a_{ {\bf{k}}\downarrow}({\bf{0}}\uparrow)
\]
\begin{equation}
+\sum_{ {\bf{k}}{\bf{q}} \neq {\bf{0}} }
(\epsilon^{e}({\bf{k}}+{\bf{q}}/2) - \epsilon^{e}({\bf{k}}-{\bf{q}}/2))
a^{\dagger}_{ {\bf{k}}\uparrow}({\bf{q}}\uparrow)
a_{ {\bf{k}}\uparrow}({\bf{q}}\uparrow)
\end{equation}
 Again it may be noted that objects such as
 $ a_{ {\bf{k}} \uparrow }({\bf{0}}\uparrow) $ and
 $ a_{ {\bf{k}} \downarrow }({\bf{0}}\downarrow) $ are omitted
 from the formalism. 
 The fact that the Fermi bilinears all evolve properly with respect to
 this hamiltonian is apparent without the need to perform any calculations.
 This is a strong indication that we are on the right track.
 It may puzzle the reader that we have included an object such as
 $ a_{ {\bf{k}}, \downarrow }({\bf{0}},\uparrow) $ in the above formula.
 This is due to the following reason.
 The commutation rule
 $ [ d_{ -{\bf{k}} }c_{ {\bf{k}} },
 c^{\dagger}_{ {\bf{k}}^{'} }d^{\dagger}_{ -{\bf{k}}^{'} }] $
 does not come out right if we don't. Let us now write down some typical
 interaction terms.
\begin{equation}
H_{e-h} = -\sum_{ {\bf{q}} \neq 0 }\frac{ v_{eh}({\bf{q}}) }{V}
\sum_{ {\bf{k}}, {\bf{k}}^{'} }
c^{\dagger}_{ {\bf{k}} + {\bf{q}}/2 }d^{\dagger}_{ -{\bf{k}}^{'} - {\bf{q}}/2 }
d_{ -{\bf{k}}^{'} + {\bf{q}}/2 }
c_{ {\bf{k}} - {\bf{q}}/2 }
\end{equation}
This may be recast in the sea-boson language as follows,
\small
\[
H_{e-h} = -\sum_{ {\bf{q}} \neq 0 }\frac{ v_{eh}({\bf{q}}) }{V}
\sum_{ {\bf{k}}, {\bf{k}}^{'} }
[\Lambda_{1}({\bf{k}}/2 + {\bf{k}}^{'}/2 + {\bf{q}}/2,{\bf{k}}^{'}
 - {\bf{k}})a^{\dagger}_{ {\bf{k}}/2 + {\bf{k}}^{'}/2 + {\bf{q}}/2
 \downarrow }({\bf{k}} - {\bf{k}}^{'} \uparrow)
 + \Lambda_{2}({\bf{k}}/2 + {\bf{k}}^{'}/2 + {\bf{q}}/2,{\bf{k}}^{'}-{\bf{k}})
a_{ {\bf{k}}/2 + {\bf{k}}^{'}/2 + {\bf{q}}/2 \uparrow }
({\bf{k}}^{'}- {\bf{k}}\downarrow)]
\]
\begin{equation}
[\Lambda_{1}({\bf{k}}/2 + {\bf{k}}^{'}/2 - {\bf{q}}/2,{\bf{k}}^{'}
 - {\bf{k}})a_{ {\bf{k}}/2 + {\bf{k}}^{'}/2
 - {\bf{q}}/2 \downarrow }({\bf{k}} - {\bf{k}}^{'} \uparrow)
 + \Lambda_{2}({\bf{k}}/2 + {\bf{k}}^{'}/2 - {\bf{q}}/2,{\bf{k}}^{'}-{\bf{k}})
a^{\dagger}_{ {\bf{k}}/2 + {\bf{k}}^{'}/2 - {\bf{q}}/2 \uparrow }
({\bf{k}}^{'}- {\bf{k}}\downarrow)]
\end{equation}
\normalsize
 Let us now try and write down the e-e/h-h repulsion terms(let us now focus
 on an undoped system),
\begin{equation}
H_{e-e} = \sum_{ {\bf{q}} \neq 0 }\frac{ v({\bf{q}}) }{2V}
\rho^{e}({\bf{q}})\rho^{e}(-{\bf{q}})
\end{equation}
\begin{equation}
H_{h-h} = \sum_{ {\bf{q}} \neq 0 }\frac{ v({\bf{q}}) }{2V}
\rho^{h}({\bf{q}})\rho^{h}(-{\bf{q}})
\end{equation}
\begin{equation}
\rho^{e}({\bf{q}}) = \sum_{ {\bf{k}} }\Lambda_{e}({\bf{k}},{\bf{q}})
a_{ {\bf{k}}\uparrow }(-{\bf{q}}\uparrow)
+ \sum_{ {\bf{k}} }
\Lambda_{e}({\bf{k}},-{\bf{q}})a^{\dagger}_{ {\bf{k}} \uparrow }
({\bf{q}}\uparrow)
\end{equation}
\begin{equation}
\rho^{h}({\bf{q}}) = -\sum_{ {\bf{k}} }\Lambda_{h}({\bf{k}},{\bf{q}})
a_{ {\bf{k}}\downarrow }(-{\bf{q}}\downarrow)
- \sum_{ {\bf{k}} }
\Lambda_{h}({\bf{k}},-{\bf{q}})a^{\dagger}_{ {\bf{k}} \downarrow }
({\bf{q}}\downarrow)
\end{equation}
 The coupling to phonons may be written as follows,
\begin{equation}
H_{ph} = \sum_{ {\bf{q}} \neq 0 }\frac{ M_{ {\bf{q}} } }{\sqrt{V}}
(b_{ {\bf{q}} } + b^{\dagger}_{ -{\bf{q}} })
(\rho^{e}({\bf{q}})-\rho^{h}({\bf{q}}))
\end{equation}
 It may be seen that only in the presence of real charge distributions do 
 the electron-electron/hole-hole repuslion and coupling to phonons contribute
 appreciably to the hamiltonian.  This means that in the undoped case in the
 absence of external fields we expect only the excitonic contribution broadened
 perhaps only via coupling to photons(which is ignored here).
 External fields, especially pump fields above the band gap cause 
 significant real populations of carriers and these in turn relax by emitting
 phonons and through Coulomb scattering. Thus the formalism we have written
 down is simple and ideal for the study of these systems.  
 The coupling to external fields may be written as,
\[
H_{ext}(t) = (\frac{|e|}{\mu c}){\vec{A}}_{ext}(t).{\vec{p}}_{vc}
\sum_{ {\bf{k}} }[\Lambda_{1}({\bf{k}},{\bf{0}})a_{ {\bf{k}}\downarrow }
({\bf{0}}\uparrow)
 + \Lambda_{2}({\bf{k}},{\bf{0}})a^{\dagger}_{ {\bf{k}}\uparrow }
({\bf{0}}\downarrow)]
\]
\begin{equation}
 +  (\frac{|e|}{\mu c}){\vec{A}}^{*}_{ext}(t).{\vec{p}}_{vc}
\sum_{ {\bf{k}} }[\Lambda_{1}({\bf{k}},{\bf{0}})a^{\dagger}
_{ {\bf{k}}\downarrow }
({\bf{0}}\uparrow)
 + \Lambda_{2}({\bf{k}},{\bf{0}})a_{ {\bf{k}}\uparrow }
({\bf{0}}\downarrow)]
\end{equation}
Let us now write down the various equations of motion of this system.
\small
\[
i\frac{ \partial }{\partial t} a_{ {\bf{k}}\downarrow }({\bf{0}}\uparrow)
 = (\epsilon^{e}({\bf{k}}) + \epsilon^{h}({\bf{k}}))
a_{ {\bf{k}}\downarrow }({\bf{0}}\uparrow)
- \sum_{ {\bf{Q}} \neq 0 }\frac{ v_{eh}({\bf{Q}}) }{V}
a_{ {\bf{k}} - {\bf{Q}}\downarrow }
({\bf{0}}\uparrow)
\]
\begin{equation}
 + \sum_{ {\bf{Q}} \neq 0 }\frac{ v_{eh}({\bf{Q}}) }{V}
[(1-\Lambda_{1}({\bf{k}}, {\bf{0}})
\Lambda_{1}({\bf{k}} - {\bf{Q}}, {\bf{0}}))
a_{ {\bf{k}} - {\bf{Q}}\downarrow }
({\bf{0}}\uparrow)
 -\Lambda_{1}({\bf{k}}, {\bf{0}})
 \Lambda_{2}({\bf{k}} - {\bf{Q}},{\bf{0}})
a^{\dagger}_{ {\bf{k}} - {\bf{Q}} \uparrow }
({\bf{0}} \downarrow)]
 + (\frac{ |e| }{\mu c}){\vec{A}}^{*}_{ext}(t).{\vec{p}}_{vc}
\Lambda_{1}({\bf{k}},{\bf{0}})
\label{FUNDEQ1}
\end{equation}
\newpage
\[
i\frac{ \partial }{\partial t} a_{ {\bf{k}}\uparrow }({\bf{0}}\downarrow)
 = -(\epsilon^{e}({\bf{k}}) + \epsilon^{h}({\bf{k}}))
a_{ {\bf{k}}\uparrow }({\bf{0}}\downarrow)
\]
\[
 - \sum_{ {\bf{Q}} \neq 0 }\frac{ v_{eh}({\bf{Q}}) }{V}
\Lambda_{2}({\bf{k}}, {\bf{0}})
[\Lambda_{1}({\bf{k}} + {\bf{Q}}, {\bf{0}})
a^{\dagger}_{ {\bf{k}} + {\bf{Q}}\downarrow }
({\bf{0}}\uparrow)
 + \Lambda_{2}({\bf{k}} + {\bf{Q}},{\bf{0}})
a_{ {\bf{k}} + {\bf{Q}} \uparrow }
({\bf{0}} \downarrow)]
\]
\begin{equation}
 + (\frac{ |e| }{\mu c}){\vec{A}}_{ext}(t).{\vec{p}}_{vc}
\Lambda_{2}({\bf{k}},{\bf{0}})
\label{FUNDEQ2}
\end{equation}
\begin{equation}
i\frac{ \partial }{\partial t}a_{ {\bf{k}} \uparrow }({\bf{q}}\uparrow)
 = \frac{ {\bf{k.q}} }{m_{e}}a_{ {\bf{k}} \uparrow }({\bf{q}}\uparrow)
 + \frac{ v({\bf{q}}) }{V}\Lambda_{e}({\bf{k}},-{\bf{q}})
\rho^{(e)}(-{\bf{q}})
 +\frac{ M_{ {\bf{q}} } }{\sqrt{V}}
X_{ {\bf{q}} }\Lambda_{e}({\bf{k}},-{\bf{q}})
\end{equation}
\begin{equation}
i\frac{ \partial }{\partial t}a_{ {\bf{k}} \downarrow }({\bf{q}}\downarrow)
 = -\frac{ {\bf{k.q}} }{m_{h}}a_{ {\bf{k}} \downarrow }({\bf{q}}\downarrow)
 - \frac{ v({\bf{q}}) }{V}\Lambda_{h}({\bf{k}},-{\bf{q}})
\rho^{(h)}(-{\bf{q}})
 + \frac{ M_{ {\bf{q}} } }{\sqrt{V}}
X_{ {\bf{q}} }\Lambda_{h}({\bf{k}},-{\bf{q}})
\end{equation}
\begin{equation}
i\frac{ \partial }{\partial t}
a^{\dagger}_{ {\bf{k}} \uparrow }(-{\bf{q}}\uparrow)
 = \frac{ {\bf{k.q}} }{m_{e}}a^{\dagger}_{ {\bf{k}} \uparrow }
(-{\bf{q}}\uparrow)
 - \frac{ v({\bf{q}}) }{V}\Lambda_{e}({\bf{k}},{\bf{q}})
\rho^{(e)}(-{\bf{q}})
 - \frac{ M_{ {\bf{q}} } }{\sqrt{V}}
X_{ {\bf{q}} }\Lambda_{e}({\bf{k}},{\bf{q}})
\end{equation}
\begin{equation}
i\frac{ \partial }{\partial t}a^{\dagger}
_{ {\bf{k}} \downarrow }(-{\bf{q}}\downarrow)
 = -\frac{ {\bf{k.q}} }{m_{h}}a^{\dagger}_{ {\bf{k}} \downarrow }
(-{\bf{q}}\downarrow)
 + \frac{ v({\bf{q}}) }{V}\Lambda_{h}({\bf{k}},{\bf{q}})
\rho^{(h)}(-{\bf{q}})
 - \frac{ M_{ {\bf{q}} } }{\sqrt{V}}
X_{ {\bf{q}} }\Lambda_{h}({\bf{k}},{\bf{q}})
\end{equation}
\begin{equation}
i\frac{ \partial }{\partial t}X_{ {\bf{q}} }
 = (2i\Omega_{LO})P_{ -{\bf{q}} }
\end{equation}
\begin{equation}
i\frac{ \partial }{\partial t}P_{ -{\bf{q}} }
 = \frac{ \Omega_{LO} }{2i}X_{ {\bf{q}} }
- i\frac{ M_{ {\bf{q}} } }{ {\sqrt{V}} }
(\rho^{(e)}(-{\bf{q}}) - \rho^{(h)}(-{\bf{q}}))
\end{equation}
\normalsize
 The last four equations of motion only affect the electron and hole 
 populations but do not impact directly upon the polarization or induced
 currents. This means that electron-electron and hole-hole repulsion 
 and electron-phonon interaction change the distributions of 
 electrons and holes and the electron-hole attraction determines the
 absorption spectrum. Let us now write down the electron and hole
 populations,
\[
{\bar{n}}_{h}({\bf{k}}) = -\langle a^{\dagger}_{ {\bf{k}} \uparrow }({\bf{0}}\downarrow)
a_{ {\bf{k}} \uparrow }({\bf{0}}\downarrow) \rangle
 + \langle a^{\dagger}_{ {\bf{k}} \downarrow }({\bf{0}}\uparrow)
a_{ {\bf{k}} \downarrow }({\bf{0}}\uparrow) \rangle
\]
\begin{equation}
-\sum_{ {\bf{q}}\neq 0 }
\langle a^{\dagger}_{ {\bf{k}}-{\bf{q}}/2 \downarrow }({\bf{q}}\downarrow)
a_{ {\bf{k}}-{\bf{q}}/2 \downarrow }({\bf{q}}\downarrow) \rangle
 + \sum_{ {\bf{q}}\neq 0 }
\langle a^{\dagger}_{ {\bf{k}}+{\bf{q}}/2  \downarrow }({\bf{q}}\downarrow)
a_{ {\bf{k}}+{\bf{q}}/2 \downarrow }({\bf{q}}\downarrow) \rangle
\label{POP1}
\end{equation}
\[
{\bar{n}}_{e}({\bf{k}}) =
 \langle a^{\dagger}_{ {\bf{k}} \downarrow }({\bf{0}}\uparrow)
a_{ {\bf{k}} \downarrow }({\bf{0}}\uparrow) \rangle
 - \langle a^{\dagger}_{ {\bf{k}} \uparrow }({\bf{0}}\downarrow)
a_{ {\bf{k}} \uparrow }({\bf{0}}\downarrow) \rangle
\]
\begin{equation}
+\sum_{ {\bf{q}}\neq 0 }
\langle a^{\dagger}_{ {\bf{k}}-{\bf{q}}/2 \uparrow }({\bf{q}}\uparrow)
a_{ {\bf{k}}-{\bf{q}}/2 \uparrow }({\bf{q}}\uparrow) \rangle
 - \sum_{ {\bf{q}}\neq 0 }
\langle a^{\dagger}_{ {\bf{k}}+{\bf{q}}/2  \uparrow }({\bf{q}}\uparrow)
a_{ {\bf{k}}+{\bf{q}}/2 \uparrow }({\bf{q}}\uparrow) \rangle
\label{POP2}
\end{equation}
 In the above sets of equations we have ignored the contribution from
 objects such as $ a_{ {\bf{k}} \uparrow }({\bf{q}}\downarrow) $
 with $ {\bf{q}} \neq 0 $. The reason being that these contributions are
 difficult to deal with. The practical consequences of this assumption
 means that we have to restric our attention to large $ {\bf{k}} $.
 Namely that we must ensure that $ {\bf{k}} $ in the above equation
 is much larger than any inverse length-scale in the problem. 
 Thus we expect our theory to be poor for $ {\bf{k}} $ small. This is in fact
 the case as we shall soon find out. The analysis including this 
 large $ {\bf{q}} $ effect 
 Let us now introduce several propagators. These are going to be useful in
 ascertaining the influence carrier-carrier repulsion and carrier-phonon
 interactions have on the populations of electrons and holes. We can see
 from Eq.(~\ref{POP1}) and Eq.(~\ref{POP2}), the manner in which the 
 populations of electrons and holes evolve. If we ignore the terms
 that involve the sea-bosons $ a_{ {\bf{k}}\sigma }({\bf{q}}\sigma) $
 thern we see that electrons and holes have the same distribution 
 determined solely by the external fields. That is, so long as relaxation
 processess are ignored we find simple and intuitively appealing 
 formulas  for the distributions. It is also worth pointing out that this
 situation is entirely analogous(perhaps even equivalent) to the
 Semiconductor Bloch Equations(SBE)\cite{Haug}.  There we find that
 if one ignores terms beyond the Hartree-Fock approximation, then
 the populations of electrons and holes are indentical even though 
 the effective masses are different. However once relaxation processes
 begin, the distributions respond appropriately and the system must be 
 solved self-consistently. A quantity such as
 $ \langle a^{\dagger}_{ {\bf{k}}\sigma }({\bf{q}}\sigma)a_{ {\bf{k}}\sigma }({\bf{q}}\sigma) \rangle $
 is zero when the electrons and holes have ''ideal''
 momentum distributions(that is, identically zero for undoped systems). 
 However, when they start to acquire non-zero values due to external 
 fields the above quantity also begins to acquire a non-zero value and the
 whole system proceeds to evolve accordingly. Let us now introduce several
 Green functions.
\begin{equation}
G_{11}({\bf{k}},{\bf{k}}^{'};{\bf{q}}) =
 -i\langle T \mbox{          }
a_{ {\bf{k}} \uparrow }({\bf{q}}\uparrow,t) 
a^{\dagger}_{ {\bf{k}}^{'} \uparrow }({\bf{q}}\uparrow,0) \rangle
\end{equation}
\begin{equation}
G_{12}({\bf{k}},{\bf{k}}^{'};{\bf{q}}) =
 -i\langle T \mbox{          }
a^{\dagger}_{ {\bf{k}} \uparrow }(-{\bf{q}}\uparrow,t)
a^{\dagger}_{ {\bf{k}}^{'} \uparrow }({\bf{q}}\uparrow,0) \rangle
\end{equation}
\begin{equation}
G_{21}({\bf{k}},{\bf{k}}^{'};{\bf{q}}) =
 -i\langle T \mbox{          }
a^{\dagger}_{ {\bf{k}} \downarrow }(-{\bf{q}}\downarrow,t)
a^{\dagger}_{ {\bf{k}}^{'} \downarrow }({\bf{q}}\downarrow,0) \rangle
\end{equation}
\begin{equation}
G_{22}({\bf{k}},{\bf{k}}^{'};{\bf{q}}) =
 -i\langle T \mbox{          }
a_{ {\bf{k}} \downarrow }({\bf{q}}\downarrow,t)
a^{\dagger}_{ {\bf{k}}^{'} \downarrow }({\bf{q}}\downarrow,0) \rangle
\end{equation}
\begin{equation}
G_{X\sigma}({\bf{k}};{\bf{q}}) = -i \langle X_{ {\bf{q}} }(t)
a^{\dagger}_{ {\bf{k}} \sigma }({\bf{q}}\sigma,0) \rangle
\end{equation}
\begin{equation}
G_{P\sigma}({\bf{k}};{\bf{q}}) = -i \langle P_{ -{\bf{q}} }(t)
a^{\dagger}_{ {\bf{k}} \sigma }({\bf{q}}\sigma,0) \rangle
\end{equation}
\[
i\frac{ \partial }{\partial t}G_{11}({\bf{k}},{\bf{k}}^{'};{\bf{q}},t)
 = \delta(t)\delta_{ {\bf{k}}, {\bf{k}}^{'} }
 + \frac{ {\bf{k.q}} }{m_{e}}
G_{11}({\bf{k}},{\bf{k}}^{'};{\bf{q}},t)
 -i \frac{ v({\bf{q}}) }{V}
\Lambda_{e}({\bf{k}},-{\bf{q}})\langle T\mbox{    }\rho^{(e)}(-{\bf{q}},t)
a^{\dagger}_{ {\bf{k}}^{'} \uparrow }({\bf{q}}\uparrow,0) \rangle
\]
\begin{equation}
+ \frac{ M_{ {\bf{q}} } }{\sqrt{V}}
\Lambda_{e}({\bf{k}},-{\bf{q}})
G_{X\uparrow}({\bf{k}}^{'};{\bf{q}},t)
\end{equation}
\[
i\frac{ \partial }{\partial t}G_{12}({\bf{k}},{\bf{k}}^{'};{\bf{q}},t)
 = \frac{ {\bf{k.q}} }{m_{e}}
G_{12}({\bf{k}},{\bf{k}}^{'};{\bf{q}},t)
 +i \frac{ v({\bf{q}}) }{V}
\Lambda_{e}({\bf{k}},{\bf{q}})\langle T\mbox{    }\rho^{(e)}(-{\bf{q}},t)
a^{\dagger}_{ {\bf{k}}^{'} \uparrow }({\bf{q}}\uparrow,0) \rangle
\]
\begin{equation}
-  \frac{ M_{ {\bf{q}} } }{\sqrt{V}}
\Lambda_{e}({\bf{k}},{\bf{q}})
G_{X\uparrow}({\bf{k}}^{'};{\bf{q}},t)
\end{equation}
\[
i\frac{ \partial }{\partial t}G_{22}({\bf{k}},{\bf{k}}^{'};{\bf{q}},t)
 = \delta(t)\delta_{ {\bf{k}}, {\bf{k}}^{'} }
 - \frac{ {\bf{k.q}} }{m_{h}}
G_{22}({\bf{k}},{\bf{k}}^{'};{\bf{q}},t)
 + i \frac{ v({\bf{q}}) }{V}
\Lambda_{h}({\bf{k}},-{\bf{q}})\langle T\mbox{    }\rho^{(h)}(-{\bf{q}},t)
a^{\dagger}_{ {\bf{k}}^{'} \downarrow }({\bf{q}}\downarrow,0) \rangle
\]
\begin{equation}
+ \frac{ M_{ {\bf{q}} } }{\sqrt{V}}
\Lambda_{h}({\bf{k}},-{\bf{q}})
G_{X\downarrow}({\bf{k}}^{'};{\bf{q}},t)
\end{equation}
\[
i\frac{ \partial }{\partial t}G_{21}({\bf{k}},{\bf{k}}^{'};{\bf{q}},t)
 = -\frac{ {\bf{k.q}} }{m_{h}}
G_{21}({\bf{k}},{\bf{k}}^{'};{\bf{q}},t)
 - i \frac{ v({\bf{q}}) }{V}
\Lambda_{h}({\bf{k}},{\bf{q}})\langle T\mbox{    }\rho^{(h)}(-{\bf{q}},t)
a^{\dagger}_{ {\bf{k}}^{'} \downarrow }({\bf{q}}\downarrow,0) \rangle
\]
\begin{equation}
- \frac{ M_{ {\bf{q}} } }{\sqrt{V}}
\Lambda_{h}({\bf{k}},{\bf{q}})
G_{X\downarrow}({\bf{k}}^{'};{\bf{q}},t)
\end{equation}
\begin{equation}
i\frac{ \partial }{\partial t}G_{X\sigma}({\bf{k}};{\bf{q}},t)
 = (2i\Omega_{LO})G_{P\sigma}({\bf{k}};{\bf{q}},t)
\end{equation}
\begin{equation}
i\frac{ \partial }{\partial t}G_{P\uparrow}({\bf{k}};{\bf{q}},t)
 = (\frac{ \Omega_{LO} }{2i})G_{X\uparrow}({\bf{k}};{\bf{q}},t)
-\frac{ M_{ {\bf{q}} } }{\sqrt{V}}
\langle T\mbox{      }\rho^{(e)}(-{\bf{q}},t)
a^{\dagger}_{ {\bf{k}} \uparrow}
({\bf{q}}\uparrow,0) \rangle
\end{equation}
\begin{equation}
i\frac{ \partial }{\partial t}G_{P\downarrow}({\bf{k}};{\bf{q}},t)
 = (\frac{ \Omega_{LO} }{2i})G_{X\downarrow}({\bf{k}};{\bf{q}},t)
+\frac{ M_{ {\bf{q}} } }{\sqrt{V}} \langle T\mbox{      }
 \rho^{(h)}(-{\bf{q}},t)a^{\dagger}_{ {\bf{k}} \downarrow}
({\bf{q}}\downarrow,0) \rangle 
\end{equation}
 In order to solve this system we have to expand the Green functions in terms
 of Matsubara frequencies.  We introduce a temperature just for 
 ease of doing calculations. In the end we shall go to the zero temperature
 limit as this is the regime when the interpretations are the cleanest
 (here $ z_{n} = 2\pi n/\beta $ and $ \beta = 1/k_{B}T $ ).
\begin{equation}
(iz_{n}-\frac{ {\bf{k.q}} }{m_{e}})
G_{11}({\bf{k}},{\bf{k}}^{'};{\bf{q}},z_{n})
 = \frac{1}{-i\beta}\delta_{ {\bf{k}}, {\bf{k}}^{'} }
 -i \frac{ {\tilde{v}}({\bf{q}},z_{n}) }{V}
\Lambda_{e}({\bf{k}},-{\bf{q}})\langle T\mbox{    }\rho^{(e)}(-{\bf{q}},z_{n})
a^{\dagger}_{ {\bf{k}}^{'} \uparrow }({\bf{q}}\uparrow,0) \rangle
\end{equation}
\begin{equation}
(iz_{n}-\frac{ {\bf{k.q}} }{m_{e}})
G_{12}({\bf{k}},{\bf{k}}^{'};{\bf{q}},z_{n})
 = i \frac{ {\tilde{v}}({\bf{q}},z_{n}) }{V}
\Lambda_{e}({\bf{k}},{\bf{q}})\langle T\mbox{    }\rho^{(e)}(-{\bf{q}},z_{n})
a^{\dagger}_{ {\bf{k}}^{'} \uparrow }({\bf{q}}\uparrow,0) \rangle
\end{equation}
\begin{equation}
(iz_{n} + \frac{ {\bf{k.q}} }{m_{h}})
G_{22}({\bf{k}},{\bf{k}}^{'};{\bf{q}},z_{n})
 = \frac{1}{-i\beta}
\delta_{ {\bf{k}}, {\bf{k}}^{'} }
 + i \frac{ {\tilde{v}}({\bf{q}}) }{V}
\Lambda_{h}({\bf{k}},-{\bf{q}})\langle T\mbox{    }\rho^{(h)}(-{\bf{q}},z_{n})
a^{\dagger}_{ {\bf{k}}^{'} \downarrow }({\bf{q}}\downarrow,0) \rangle
\end{equation}
\begin{equation}
(iz_{n} + \frac{ {\bf{k.q}} }{m_{h}})
G_{21}({\bf{k}},{\bf{k}}^{'};{\bf{q}},z_{n})
 = -i \frac{ {\tilde{v}}({\bf{q}},z_{n}) }{V}
\Lambda_{h}({\bf{k}},{\bf{q}})\langle T\mbox{    }\rho^{(h)}(-{\bf{q}},z_{n})
a^{\dagger}_{ {\bf{k}}^{'} \downarrow }({\bf{q}}\downarrow,0) \rangle
\end{equation}
\begin{equation}
{\tilde{v}}({\bf{q}},z_{n}) = v({\bf{q}}) 
-\frac{ 2\mbox{        }
M^{2}_{ {\bf{q}} }\Omega_{LO} }{ z^{2}_{n} + \Omega^{2}_{LO} }
\end{equation}
\begin{equation}
-i\langle T\mbox{         }\rho^{(e)}(-{\bf{q}},z_{n})
a^{\dagger}_{ {\bf{k}}^{'}\uparrow }({\bf{q}}\uparrow,0)\rangle
 = \sum_{ {\bf{p}} }\Lambda_{e}({\bf{p}},-{\bf{q}})
G_{11}({\bf{p}},{\bf{k}}^{'};{\bf{q}},z_{n})
 + \sum_{ {\bf{p}} }\Lambda_{e}({\bf{p}},{\bf{q}})
G_{12}({\bf{p}},{\bf{k}}^{'};{\bf{q}},z_{n})
\end{equation}
\begin{equation}
-i\langle T\mbox{         }\rho^{(h)}(-{\bf{q}},z_{n})
a^{\dagger}_{ {\bf{k}}^{'}\downarrow }({\bf{q}}\downarrow,0)\rangle
 = -\sum_{ {\bf{p}} }\Lambda_{h}({\bf{p}},-{\bf{q}})
G_{22}({\bf{p}},{\bf{k}}^{'};{\bf{q}},z_{n})
 - \sum_{ {\bf{p}} }\Lambda_{h}({\bf{p}},{\bf{q}})
G_{21}({\bf{p}},{\bf{k}}^{'};{\bf{q}},z_{n})
\end{equation}
\[
-i\langle T \mbox{          }\rho^{(e)}(-{\bf{q}},z_{n})
a^{\dagger}_{ {\bf{k}}^{'}\uparrow }({\bf{q}}\uparrow,0) \rangle
 = (\frac{1}{-i\beta})\frac{ \Lambda_{e}({\bf{k}}^{'},-{\bf{q}}) }
{iz_{n} - \frac{ {\bf{k}}^{'}.{\bf{q}} }{m_{e}}}
\]
\begin{equation}
+ i\frac{ {\tilde{v}}({\bf{q}},z_{n}) }{V}
\sum_{ {\bf{p}} } \frac{ \Lambda^{2}_{e}({\bf{p}},{\bf{q}})
 - \Lambda^{2}_{e}({\bf{p}},-{\bf{q}}) }
{iz_{n} - \frac{ {\bf{p.q}} }{m_{e}} }
\langle T \mbox{          }\rho^{(e)}(-{\bf{q}},z_{n})
a^{\dagger}_{ {\bf{k}}^{'}\uparrow }({\bf{q}}\uparrow,0) \rangle
\end{equation}
\begin{equation}
\langle T \mbox{          }\rho^{(e)}(-{\bf{q}},z_{n})
a^{\dagger}_{ {\bf{k}}^{'}\uparrow }({\bf{q}}\uparrow,0) \rangle
 = (\frac{1}{-i\beta})\frac{ i\Lambda_{e}({\bf{k}}^{'},-{\bf{q}}) }
{iz_{n} - \frac{ {\bf{k}}^{'}.{\bf{q}} }{m_{e}}}
\frac{1}{\epsilon^{(e)}({\bf{q}},iz_{n}) }
\end{equation}
where,
\begin{equation}
\epsilon^{(e)}({\bf{q}},iz_{n}) = 1 + 
\frac{ {\tilde{v}}({\bf{q}},z_{n}) }{V}
\sum_{ {\bf{p}} }\frac{ {\bar{n}}^{(e)}({\bf{p}}+{\bf{q}}/2)
 - {\bar{n}}^{(e)}({\bf{p}}-{\bf{q}}/2) }
{iz_{n} - \frac{ {\bf{p.q}} }{m_{e}}}
\end{equation}
\begin{equation}
\langle T \mbox{          }\rho^{(h)}(-{\bf{q}},z_{n})
a^{\dagger}_{ {\bf{k}}^{'}\downarrow }({\bf{q}}\downarrow,0) \rangle
 = (\frac{1}{-i\beta})\frac{ -i\Lambda_{h}({\bf{k}}^{'},-{\bf{q}}) }
{iz_{n} + \frac{ {\bf{k}}^{'}.{\bf{q}} }{m_{h}}}
\frac{1}{\epsilon^{(h)}({\bf{q}},iz_{n})}
\end{equation}
where,
\begin{equation}
\epsilon^{(h)}({\bf{q}},iz_{n}) = 1 +
\frac{ {\tilde{v}}({\bf{q}},z_{n}) }{V}
\sum_{ {\bf{p}} }\frac{ {\bar{n}}^{(h)}({\bf{p}}+{\bf{q}}/2)
 - {\bar{n}}^{(h)}({\bf{p}}-{\bf{q}}/2) }
{ iz_{n} - \frac{ {\bf{p.q}} }{m_{h}} }
\end{equation}
Therefore,
\begin{equation}
\langle a^{\dagger}_{ {\bf{k}} \uparrow }({\bf{q}}\uparrow)
a_{ {\bf{k}} \uparrow }({\bf{q}}\uparrow) \rangle
 = (\frac{1}{-i\beta})(\frac{1}{V})
\sum_{n}
\frac{ {\tilde{v}}({\bf{q}},z_{n}) }{ \epsilon^{(e)}({\bf{q}},iz_{n}) }
\frac{ i\Lambda^{2}_{e}({\bf{k}},-{\bf{q}}) }
{ (iz_{n} - \frac{ {\bf{k.q}} }{m_{e}})^{2} }
\end{equation}
\begin{equation}
\langle a^{\dagger}_{ {\bf{k}} \downarrow }({\bf{q}}\downarrow)
a_{ {\bf{k}} \downarrow }({\bf{q}}\downarrow) \rangle
 = (\frac{1}{-i\beta})(\frac{1}{V})
\sum_{n}
\frac{ {\tilde{v}}({\bf{q}},z_{n}) }{ \epsilon^{(h)}({\bf{q}},iz_{n}) }
\frac{ i\Lambda^{2}_{h}({\bf{k}},-{\bf{q}}) }
{ (iz_{n} + \frac{ {\bf{k.q}} }{m_{h}})^{2} }
\end{equation}
 Then when we go to the zero-temperature limit we have
 to integrate over all $ n $.  In Fig.1 we see the pole structure of the
 above equations. Call $ i\mbox{       }z_{n} = i\mbox{        }z $. Then,
\begin{equation}
\langle a^{\dagger}_{ {\bf{k}} \uparrow }({\bf{q}}\uparrow)
a_{ {\bf{k}} \uparrow }({\bf{q}}\uparrow) \rangle
 =
(\frac{1}{-i\beta})(\frac{1}{V})
\frac{\beta}{2\pi}\int_{C} dz\mbox{     }
\frac{ {\tilde{v}}({\bf{q}},z) }{ \epsilon^{(e)}({\bf{q}},iz) }
\frac{ i\Lambda^{2}_{e}({\bf{k}},-{\bf{q}}) }
{ (iz - \frac{ {\bf{k.q}} }{m_{e}})^{2} }
\end{equation}
 In Fig.1 we see the pole structure of the above contour integral. 
 Let us assume that $ {\bf{k.q}} > 0 $ then the pole 
 $ z = -i \mbox{      }{\bf{k.q}}/m_{e} $ is in the lower half-plane. 
 We have to close the contour in such a way that this pole is excluded
 from consideration, since if it were included we would have a formula for
 $ \langle a_{ {\bf{k}}\uparrow }({\bf{q}}\uparrow) a^{\dagger}_{ {\bf{k}}\uparrow }({\bf{q}}\uparrow) \rangle $
 rather than $ \langle a^{\dagger}_{ {\bf{k}} \uparrow }({\bf{q}}\uparrow)a_{ {\bf{k}} \uparrow }({\bf{q}}\uparrow) \rangle $.
 Therefore we have to close the contuor in the upper half-plane(and
 $ C = C1 $). 
 Now, if we count the number of poles in the integrand we find
 that first of all, the zeros of the dielectric function
 that lie on the
 positive imaginary axis of the z-plane(how many zeros are there, is an
 important question which we shall address subsequently) contribute.
 Then it seems at first sight that even the poles of
 $ {\tilde{v}}({\bf{q}},z) $
 contribute. The poles of this function lie at
 $ \pm i\mbox{  }\Omega_{LO} $. However, upon closer examination we find that
 this is not the case. The poles of $ {\tilde{v}}({\bf{q}},z) $ are
 also poles of $  \epsilon^{(e)}({\bf{q}},iz) $ and the two cancel. 
 Thus the only poles that contribute are the zeros of the dielectric function 
 that lie on the positive imaginary axis. How many such zeros are there ?
 If one counts only the
 collective modes then one arrives at the conclusion that there are only
 two, one corresponding to the plasmon(modified by phonons) and the other
 corresponding to phonons(modified by Coulomb interactions). There is another
 mode that is equally important, indeed it would be a serious mistake to
 ignore this contribution, namely the particle-hole mode. We have encountered
 this problem before\cite{Setlur}. In our earlier article we presented an
 argument\cite{Setlur}
 that shows how one may incorporate the particle-hole mode. In retrospect it
 seems that the approach presented there is not a good one, although it 
 serves well to illustrate the importance of the particle-hole mode.
 Here we shall take the point of view that all energies are allowed as 
 zeros of the dielectric function(for each $ {\bf{q}} $) but each
 comes with a weight corresponding to the strength of the dynamical structure
 factor at that energy. Thus for small $ {\bf{q}} $ we recover naturally
 the collective modes but for larger $ {\bf{q}} $ we start summing the
 particle-hole modes as well. There is really is no rigorous justification
 for this point of view except that it is physically well-motivated. 
\begin{equation}
\langle a^{\dagger}_{ {\bf{k}} \uparrow }({\bf{q}}\uparrow)
a_{ {\bf{k}} \uparrow }({\bf{q}}\uparrow) \rangle
 =
(\frac{1}{i})(\frac{1}{V})
\int_{C} \frac{dz}{2\pi\mbox{       }i}\mbox{     }
\frac{ {\tilde{v}}({\bf{q}},z) }{ \epsilon^{(e)}({\bf{q}},iz) }
\frac{ \Lambda^{2}_{e}({\bf{k}},-{\bf{q}}) }
{ (iz - \frac{ {\bf{k.q}} }{m_{e}})^{2} }
\end{equation}
The poles are $ z = i\mbox{         }\omega^{(e)}_{I}({\bf{q}}) $,
 $ \omega_{I}^{(e)}({\bf{q}}) > 0$ satisfies
 $ \epsilon^{(e)}({\bf{q}}, \omega^{(e)}_{I}) = 0 $.
\begin{equation}
\langle a^{\dagger}_{ {\bf{k}} \uparrow }({\bf{q}}\uparrow)
a_{ {\bf{k}} \uparrow }({\bf{q}}\uparrow) \rangle
 =
(\frac{1}{i})(\frac{1}{V})
\sum_{I} 
 \frac{ {\tilde{v}}({\bf{q}},i\mbox{         }\omega^{(e)}_{I}) }
{ \frac{ \partial }{\partial z}|_{ z = i\mbox{         }
\omega^{(e)}_{I}({\bf{q}}) }
\epsilon^{(e)}({\bf{q}},iz) }
\frac{ \Lambda^{2}_{e}({\bf{k}},-{\bf{q}}) }
{ (\omega^{(e)}_{I}({\bf{q}}) +  \frac{ {\bf{k.q}} }{m_{e}})^{2} }
\label{ELECEQN1}
\end{equation}
Let us first evaluate this.   
Similarly, one may write for holes,
\begin{equation}
\langle a^{\dagger}_{ {\bf{k}} \downarrow }({\bf{q}}\downarrow)
a_{ {\bf{k}} \downarrow }({\bf{q}}\downarrow) \rangle
 =
(\frac{1}{i})(\frac{1}{V})
\sum_{I} 
 \frac{ {\tilde{v}}({\bf{q}},i\mbox{         }\omega^{(h)}_{I}) }
{ \frac{ \partial }{\partial z}|_{ z = i\mbox{         }
\omega^{(h)}_{I}({\bf{q}}) }
\epsilon^{(h)}({\bf{q}},iz) }
\frac{ \Lambda^{2}_{h}({\bf{k}},-{\bf{q}}) }
{ (\omega^{(h)}_{I}({\bf{q}}) - \frac{ {\bf{k.q}} }{m_{h}})^{2} }
\label{HOLEEQN1}
\end{equation}
 Let us now evaluate these quantities more explicitly.
\[
\frac{ \partial }{ \partial z }|_{z = i\omega_{I} }
\epsilon({\bf{q}},iz)
 = -P({\bf{q}},iz)|_{ z = i\omega_{I} }
\frac{ \partial }{ \partial z }|_{z = i\omega_{I} }
 {\tilde{v}}({\bf{q}},z)
 - {\tilde{v}}({\bf{q}},z)|_{z = i\omega_{I} }
\frac{ \partial }{ \partial z }|_{z = i\omega_{I} }P({\bf{q}},iz)
\]
\[
 = -\frac{1}{ {\tilde{v}}({\bf{q}},z) }|_{z = i\omega_{I} }
\frac{ \partial }{ \partial z }|_{z = i\omega_{I} }
 {\tilde{v}}({\bf{q}},z)
 - {\tilde{v}}({\bf{q}},z)|_{z = i\omega_{I} }
\frac{ \partial }{ \partial z }|_{z = i\omega_{I} }P({\bf{q}},iz)
\]
Since,
\begin{equation}
{\tilde{v}}({\bf{q}},z)
 = v({\bf{q}}) - \frac{ 2 \Omega_{LO} M^{2}_{ {\bf{q}} } }
{ z^{2} + \Omega^{2}_{LO} }
\end{equation}
\begin{equation}
\frac{ \partial }{ \partial z }|_{z = i\omega_{I} }
P({\bf{q}},iz) = \frac{i}{V}\sum_{ {\bf{k}} }
\frac{ {\bar{n}}_{ {\bf{k}} - {\bf{q}}/2 }
 - {\bar{n}}_{ {\bf{k}} + {\bf{q}}/2 } }
{( \omega_{I} - \frac{ {\bf{k.q}} }{m})^{2} }
\end{equation}
\begin{equation}
\frac{ \partial }{ \partial z }|_{z = i\omega_{I} }{\tilde{v}}({\bf{q}},z) = 
 \frac{ 4 i\omega_{I}\Omega_{LO} M^{2}_{ {\bf{q}} } }
{ (\omega^{2}_{I} -  \Omega^{2}_{LO})^{2} }
\end{equation}
\begin{equation}
\frac{ \partial }{ \partial z }|_{z = i\omega_{I} }\epsilon({\bf{q}},iz)
 = -i \frac{ V }{ {\tilde{v}}({\bf{q}},i\omega_{I})}
\{\frac{ M^{2}_{ {\bf{q}} } }{V} \frac{ 4\Omega_{LO} \omega_{I} } 
{(\omega^{2}_{I} - \Omega^{2}_{LO})^{2}}
+ (\frac{ {\tilde{v}}({\bf{q}},i\omega_{I}) }{V})^{2}
\sum_{ {\bf{k}} } \frac{ {\bar{n}}_{ {\bf{k}} - {\bf{q}}/2 }
 - {\bar{n}}_{ {\bf{k}} + {\bf{q}}/2 } }
{ ( \omega_{I} - \frac{ {\bf{k.q}} }{m})^{2} }\}
\end{equation}
 As we pointed our just a while ago, it is necessary that we interpret
 the sum over $ I $ in a special manner so that we
 are able to recover both the collective as well as the
 particle-hole modes. The way this is done is through the following
 identification,
\begin{equation}
\sum_{ I,{\bf{q}} }f({\bf{q}}, \omega_{I}) = 
\sum_{ {\bf{q}} }\int_{0}^{\infty} d\omega\mbox{      }
W({\bf{q}},\omega)f({\bf{q}}, \omega)
\end{equation}
where the weight is the dynamical structure factor normalised to unity.
\begin{equation}
W({\bf{q}},\omega) = \frac{ S({\bf{q}},\omega) }
{ \int_{0}^{\infty} d\omega \mbox{      } S({\bf{q}},\omega) }
\end{equation}
 The dynamical structure factor is defined to be the the dynamical
 density-density correlation function fourier-transformed 
 divided by the total number of particles. Let us first write down,
\begin{equation}
  \langle T \mbox{          }\rho^{(e)}(-{\bf{q}},t)
a^{\dagger}_{ {\bf{k}}^{'}\uparrow }({\bf{q}}\uparrow,0) \rangle
 = (\frac{1}{-i\beta})
\sum_{n} e^{z_{n}t}
\frac{ i\Lambda_{e}({\bf{k}}^{'},-{\bf{q}}) }
{iz_{n} - \frac{ {\bf{k}}^{'}.{\bf{q}} }{m_{e}}}
\frac{1}{\epsilon^{(e)}({\bf{q}},iz_{n}) }
\end{equation}
Since $ Im(t) \in [0, -\beta] $, if $ Im(t) < 0 $ then,
\[
  \langle \rho^{(e)}(-{\bf{q}},t)
a^{\dagger}_{ {\bf{k}}^{'}\uparrow }({\bf{q}}\uparrow,0) \rangle
 = (\frac{1}{-i\beta})(\frac{\beta }{2 \pi})
\int_{C_{-}} \mbox{        }dz \mbox{       } 
e^{zt}\mbox{       }
\frac{ i\Lambda_{e}({\bf{k}}^{'},-{\bf{q}}) }
{ iz - \frac{ {\bf{k}}^{'}.{\bf{q}} }{m_{e}} }
\frac{1}{ \epsilon^{(e)}({\bf{q}},iz) }
\]
\begin{equation}
= (\frac{i}{2\pi})\int^{+\infty}_{-\infty}dx\mbox{        }
e^{xt}\mbox{      }\frac{ i\Lambda_{e}({\bf{k}}^{'},-{\bf{q}}) }
{ix - \frac{ {\bf{k}}^{'}.{\bf{q}} }{m_{e}} }
\frac{1}{\epsilon^{(e)}({\bf{q}},ix) }
\end{equation}
If $ Im(t) > 0 $ then,
\[
  \langle a^{\dagger}_{ {\bf{k}}^{'}\uparrow }({\bf{q}}\uparrow,0)
\rho^{(e)}(-{\bf{q}},t) \rangle
 = (\frac{1}{-i\beta})(\frac{\beta }{2 \pi})
\int_{C_{+}} \mbox{        }dz \mbox{       } 
e^{zt}\mbox{         }
\frac{ i\Lambda_{e}({\bf{k}}^{'},-{\bf{q}}) }
{iz - \frac{ {\bf{k}}^{'}.{\bf{q}} }{m_{e}}}
\frac{1}{\epsilon^{(e)}({\bf{q}},iz) }
\]
\begin{equation}
  = (\frac{i }{2 \pi})
\int^{+\infty}_{-\infty} \mbox{        }dx \mbox{       } 
e^{xt}\mbox{         }
\frac{ i\Lambda_{e}({\bf{k}}^{'},-{\bf{q}}) }
{ix - \frac{ {\bf{k}}^{'}.{\bf{q}} }{m_{e}} }
\frac{1}{\epsilon^{(e)}({\bf{q}},ix) }
\end{equation}
 Here $ C_{+} (C_{-}) $ is the semi-circle in the upper(lower) half plane.  
 Let us now take the complex conjugate of the above equation.
 $ Im(t) > 0 $ implies $ Im(t^{*}) < 0 $. Therefore if $ Im(t^{*}) < 0 $,
\begin{equation}
  \langle  \rho^{(e)}({\bf{q}},t^{*})
 a_{ {\bf{k}}^{'}\uparrow }({\bf{q}}\uparrow,0) \rangle
  = -(\frac{i }{2 \pi})
\int^{+\infty}_{-\infty} \mbox{        }dx \mbox{       } 
e^{xt^{*}}\mbox{         }
\frac{ -i\Lambda_{e}({\bf{k}}^{'},-{\bf{q}}) }
{-ix - \frac{ {\bf{k}}^{'}.{\bf{q}} }{m_{e}} }
\frac{1}{\epsilon^{(e)}({\bf{q}},ix) }
\end{equation}
Define,
\begin{equation}
\rho^{(e,a)}({\bf{q}},0) = \sum_{ {\bf{k}}^{'} }
\Lambda_{e}({\bf{k}}^{'},-{\bf{q}})a^{\dagger}_{ {\bf{k}}^{'} \uparrow }
({\bf{q}}\uparrow,0) 
\end{equation}
\begin{equation}
\rho^{(e,b)}({\bf{q}},0) = \sum_{ {\bf{k}}^{'} }
\Lambda_{e}({\bf{k}}^{'},{\bf{q}})a_{ {\bf{k}}^{'} \uparrow }
(-{\bf{q}}\uparrow,0) 
\end{equation}
therefore,
\begin{equation}
\rho^{(e)}({\bf{q}},0) = \rho^{(e,a)}({\bf{q}},0) + \rho^{(e,b)}({\bf{q}},0)
\end{equation}
Then we have $ Im(t) < 0 $,
\begin{equation}
\langle \rho^{(e)}(-{\bf{q}},t)\rho^{(e)}({\bf{q}},0) \rangle
 =  (\frac{i}{2\pi})\int^{+\infty}_{-\infty}dx\mbox{        }
e^{xt}\mbox{      }
\sum_{ {\bf{k}}^{'} }
\frac{ i ({\bar{n}}_{e}({\bf{k}}^{'}-{\bf{q}}/2)
 - {\bar{n}}_{e}({\bf{k}}^{'}+{\bf{q}}/2)) }
{ix - \frac{ {\bf{k}}^{'}.{\bf{q}} }{m_{e}} }
\frac{1}{ \epsilon^{(e)}({\bf{q}},ix) }
\end{equation}
For $ Im(t^{*}) > 0 $,
\begin{equation}
\langle \rho^{(e)}({\bf{q}},0)\rho^{(e)}(-{\bf{q}},t^{*}) \rangle
 =  (\frac{i}{2\pi})\int^{+\infty}_{-\infty}dx\mbox{        }
e^{xt^{*}}\mbox{      }
\sum_{ {\bf{k}}^{'} }
\frac{ i ({\bar{n}}_{e}({\bf{k}}^{'}+{\bf{q}}/2)
 - {\bar{n}}_{e}({\bf{k}}^{'}-{\bf{q}}/2)) }
{-ix + \frac{ {\bf{k}}^{'}.{\bf{q}} }{m_{e}} }
\frac{1}{ \epsilon^{(e)}({\bf{q}},ix) }
\end{equation}
Combining these two,
\[
\langle T\rho^{(e)}(-{\bf{q}},t)\rho^{(e)}({\bf{q}},0) \rangle
 =  (\frac{i}{2\pi})\int^{+\infty}_{-\infty}dx\mbox{        }
e^{xt}\mbox{      }
\sum_{ {\bf{k}}^{'} }
\frac{ i ({\bar{n}}_{e}({\bf{k}}^{'}-{\bf{q}}/2)
 - {\bar{n}}_{e}({\bf{k}}^{'}+{\bf{q}}/2)) }
{ix - \frac{ {\bf{k}}^{'}.{\bf{q}} }{m_{e}} }
\frac{1}{ \epsilon^{(e)}({\bf{q}},ix) }
\]
\begin{equation}
 = -(\frac{V}{2\pi})\int^{+\infty}_{-\infty}dx\mbox{        }
e^{xt}\mbox{      }\frac{ P_{e}({\bf{q}},ix) }{ \epsilon^{(e)}({\bf{q}},ix) }
\end{equation} 
Define the Green function\cite{Mahan},
\begin{equation}
{\mathcal{D}}({\bf{q}},t) = -\langle T\rho^{(e)}(-{\bf{q}},t)
\rho^{(e)}({\bf{q}},0) \rangle
\end{equation}
then,
\begin{equation}
{\mathcal{D}}_{ret}({\bf{q}},\omega) = 
V\frac{ P^{ret}_{e}({\bf{q}},\omega) }{ \epsilon_{ret}^{(e)}({\bf{q}},\omega) }
\end{equation}
The corresponding spectral function is the dynamical structure factor,
\begin{equation}
 N_{e}\mbox{       }
S({\bf{q}},\omega) = -2 Im( {\mathcal{D}}_{ret}({\bf{q}},\omega) )
\end{equation}
\begin{equation}
 \epsilon_{r}^{(e)}({\bf{q}},\omega) =  1 - v^{r}({\bf{q}},\omega)
P^{r}_{e}({\bf{q}},\omega) + v^{i}({\bf{q}},\omega)P^{i}_{e}({\bf{q}},\omega)
\end{equation}
\begin{equation}
 \epsilon_{i}^{(e)}({\bf{q}},\omega) =   - v^{r}({\bf{q}},\omega)
P^{i}_{e}({\bf{q}},\omega) - v^{i}({\bf{q}},\omega)P^{r}_{e}({\bf{q}},\omega)
\end{equation} 
 This procedure ensures that we correctly incorporate both the collective
 (for small $ {\bf{q}} $) and the particle-hole modes.
 There is an alternative approach that comes to mind. That is the method of
 exact diagonalisation.  Consider the hamiltonian,
\[
H^{'} = \sum_{ {\bf{k}},{\bf{q}} }\frac{ {\bf{k.q}} }{m_{e}}
a^{\dagger}_{ {\bf{k}}\uparrow }({\bf{q}}\uparrow)
a_{ {\bf{k}}\uparrow }({\bf{q}}\uparrow)
 - \sum_{ {\bf{k}},{\bf{q}} }\frac{ {\bf{k.q}} }{m_{h}}
a^{\dagger}_{ {\bf{k}}\downarrow }({\bf{q}}\downarrow)
a_{ {\bf{k}}\downarrow }({\bf{q}}\downarrow)
+ \sum_{ {\bf{q}} }\Omega_{LO} b^{\dagger}_{ {\bf{q}} }b_{ {\bf{q}} }
\]
\[
+ \sum_{ {\bf{q}} \neq {\bf{0}} }\frac{ v({\bf{q}}) }{2V}
\sum_{ {\bf{k}}, {\bf{k}}^{'} }
[\Lambda_{e}({\bf{k}},{\bf{q}})a_{ {\bf{k}} \uparrow }(-{\bf{q}}\uparrow)
 + \Lambda_{e}({\bf{k}},-{\bf{q}})a^{\dagger}_{ {\bf{k}} \uparrow }
({\bf{q}}\uparrow)]
[\Lambda_{e}({\bf{k}}^{'},-{\bf{q}})a_{ {\bf{k}}^{'} \uparrow }
({\bf{q}}\uparrow)
 + \Lambda_{e}({\bf{k}}^{'},{\bf{q}})a^{\dagger}_{ {\bf{k}}^{'} \uparrow }
(-{\bf{q}}\uparrow)]
\]
\[
+ \sum_{ {\bf{q}} \neq {\bf{0}} }\frac{ v({\bf{q}}) }{2V}
\sum_{ {\bf{k}}, {\bf{k}}^{'} }
[\Lambda_{h}({\bf{k}},{\bf{q}})a_{ {\bf{k}} \downarrow }(-{\bf{q}}\downarrow)
 + \Lambda_{h}({\bf{k}},-{\bf{q}})a^{\dagger}_{ {\bf{k}} \downarrow }
({\bf{q}}\downarrow)]
[\Lambda_{h}({\bf{k}}^{'},-{\bf{q}})a_{ {\bf{k}}^{'} \downarrow }
({\bf{q}}\downarrow)
 + \Lambda_{h}({\bf{k}}^{'},{\bf{q}})a^{\dagger}_{ {\bf{k}}^{'} \downarrow }
(-{\bf{q}}\downarrow)]
\]
\begin{equation}
+ \sum_{ {\bf{q}} \neq {\bf{0}} }\frac{ M_{ {\bf{q}} } }{ \sqrt{V} }
(b_{ {\bf{q}} } + b^{\dagger}_{ -{\bf{q}} })
[\Lambda_{e}({\bf{k}},{\bf{q}})a_{ {\bf{k}} \uparrow }(-{\bf{q}}\uparrow)
 + \Lambda_{e}({\bf{k}},-{\bf{q}})a^{\dagger}_{ {\bf{k}} \uparrow }
({\bf{q}}\uparrow)
 + \Lambda_{h}({\bf{k}},{\bf{q}})a_{ {\bf{k}} \downarrow }(-{\bf{q}}\downarrow)
 + \Lambda_{h}({\bf{k}},-{\bf{q}})a^{\dagger}_{ {\bf{k}} \downarrow }
({\bf{q}}\downarrow)]
\end{equation}
 In order to diagonalise this we proceed as follows. Let us postulate the
 existence of dressesd sea-bosons $ d_{I\sigma}({\bf{q}}) $ such that,
\begin{equation}
H^{'} = \sum_{I,{\bf{q}},\sigma }\omega_{I\sigma}({\bf{q}})
d^{\dagger}_{I\sigma}({\bf{q}})d_{I\sigma}({\bf{q}})
\end{equation}
 Now for some notation. $ \sigma = e, h $
 (correspondingly $ \sigma = \uparrow, \downarrow $, furthermore,
 $ S(\sigma) $ is such that $ S(\uparrow) =  +1 $ and
 $ S(\downarrow) =  -1 $).  
\begin{equation}
a_{ {\bf{k}}\sigma }({\bf{q}}\sigma) 
 = \sum_{I} [a_{ {\bf{k}}\sigma }({\bf{q}}\sigma),
d^{\dagger}_{I\sigma}({\bf{q}})]d_{I\sigma}({\bf{q}})
 - \sum_{I} [a_{ {\bf{k}}\sigma }({\bf{q}}\sigma),
d_{I\sigma}(-{\bf{q}})]d^{\dagger}_{I\sigma}(-{\bf{q}})
\end{equation}
\begin{equation}
b_{ {\bf{q}} } = \sum_{I,\sigma}
[b_{ {\bf{q}} }, d^{\dagger}_{I\sigma}({\bf{q}})]d_{I\sigma}({\bf{q}})
 - \sum_{I,\sigma}
[b_{ {\bf{q}} }, d_{I\sigma}(-{\bf{q}})]d^{\dagger}_{I\sigma}(-{\bf{q}})
\end{equation}
The inverse relation is,
\[
d_{I\sigma}({\bf{q}}) 
 = \sum_{ {\bf{k}} } [d_{I\sigma}({\bf{q}}),
a^{\dagger}_{ {\bf{k}}\sigma }({\bf{q}}\sigma)]
a_{ {\bf{k}}\sigma }({\bf{q}}\sigma)
 - \sum_{ {\bf{k}} } [d_{I\sigma}({\bf{q}}),
a_{ {\bf{k}}\sigma }(-{\bf{q}}\sigma)]
a^{\dagger}_{ {\bf{k}}\sigma }(-{\bf{q}}\sigma)
\]
\begin{equation}
 + [d_{I\sigma}({\bf{q}}), b^{\dagger}_{ {\bf{q}} }]b_{ {\bf{q}} }
 - [d_{I\sigma}({\bf{q}}), b_{ -{\bf{q}} }]b^{\dagger}_{ -{\bf{q}} }
\end{equation}
Therefore,
\[
\omega_{I\sigma}({\bf{q}})d_{I\sigma}({\bf{q}}) = \sum_{ {\bf{k}} }
S(\sigma)\frac{ {\bf{k.q}} }{m_{\sigma}}[d_{I\sigma}({\bf{q}}),
 a^{\dagger}_{ {\bf{k}} \sigma }({\bf{q}}\sigma) ]
a_{ {\bf{k}} \sigma }({\bf{q}}\sigma) 
 - S(\sigma)
\sum_{ {\bf{k}} }\frac{ {\bf{k.q}} }{m_{\sigma}}[d_{I\sigma}({\bf{q}}),
 a_{ {\bf{k}} \sigma }(-{\bf{q}}\sigma) ]
a^{\dagger}_{ {\bf{k}} \sigma }(-{\bf{q}}\sigma)
\]
\[
+ \frac{ v({\bf{q}}) }{V}\sum_{ {\bf{k}}, {\bf{k}}^{'} }
[\Lambda_{\sigma}({\bf{k}},{\bf{q}})
[d_{I\sigma}({\bf{q}}),a_{ {\bf{k}}\sigma}(-{\bf{q}}\sigma)]
 + \Lambda_{\sigma}({\bf{k}},-{\bf{q}})
[d_{I\sigma}({\bf{q}}),a^{\dagger}_{ {\bf{k}}\sigma}({\bf{q}}\sigma)]]
[\Lambda_{\sigma}({\bf{k}}^{'},-{\bf{q}})a_{ {\bf{k}}^{'} \sigma }
({\bf{q}}\sigma)
 + \Lambda_{\sigma}({\bf{k}}^{'},{\bf{q}})a^{\dagger}_{ {\bf{k}}^{'}\sigma}
(-{\bf{q}}\sigma)]
\]
\[
 + \Omega_{LO}[d_{I\sigma}({\bf{q}}),b^{\dagger}_{ {\bf{q}} }]b_{ {\bf{q}} }
 + \Omega_{LO}[d_{I\sigma}({\bf{q}}),b_{ -{\bf{q}} }]b^{\dagger}_{ -{\bf{q}} }
\]
\[
+ \frac{ M_{ {\bf{q}} } }{\sqrt{V}}([d_{I\sigma}({\bf{q}}),b_{ -{\bf{q}} }]
 + [d_{I\sigma}({\bf{q}}),b^{\dagger}_{ {\bf{q}} }]) 
\sum_{ {\bf{k}} }[\Lambda_{\sigma}({\bf{k}},-{\bf{q}})
a_{ {\bf{k}}\sigma }({\bf{q}}\sigma) 
 + \Lambda_{\sigma}({\bf{k}},{\bf{q}})
a^{\dagger}_{ {\bf{k}}\sigma }(-{\bf{q}}\sigma)]
\]
\begin{equation}
+ \frac{ M_{ {\bf{q}} } }{\sqrt{V}}(b_{ {\bf{q}} }+b^{\dagger}_{ -{\bf{q}} })
\sum_{ {\bf{k}} }(\Lambda_{\sigma}({\bf{k}},{\bf{q}})
[d_{I\sigma}({\bf{q}}),a_{ {\bf{k}}\sigma }(-{\bf{q}}\sigma)]
 + \Lambda_{\sigma}({\bf{k}},-{\bf{q}})
[d_{I\sigma}({\bf{q}}),a^{\dagger}_{ {\bf{k}}\sigma }({\bf{q}}\sigma)]
\end{equation}
\begin{equation}
(\omega_{I\sigma}({\bf{q}})-S(\sigma)\frac{ {\bf{k.q}} }{m_{\sigma}})
[d_{I\sigma}({\bf{q}}),
a^{\dagger}_{ {\bf{k}}\sigma }({\bf{q}}\sigma)]
 = S(\sigma)\frac{ {\tilde{v}}({\bf{q}},I\sigma) }{V}\Lambda_{\sigma}
({\bf{k}},-{\bf{q}})
\rho^{(\sigma)}({\bf{q}},I)
\end{equation}
\begin{equation}
(\omega_{I\sigma}({\bf{q}})-S(\sigma)\frac{ {\bf{k.q}} }{m_{\sigma}})
[d_{I\sigma}({\bf{q}}),a_{ {\bf{k}}\sigma }(-{\bf{q}}\sigma)]
 = -S(\sigma)
\frac{ {\tilde{v}}({\bf{q}},I\sigma) }{V}\Lambda_{\sigma}({\bf{k}},{\bf{q}})
\rho^{(\sigma)}({\bf{q}},I)
\end{equation}
\begin{equation}
\rho^{(e)}({\bf{q}},I) = \sum_{ {\bf{k}} }\Lambda_{e}({\bf{k}},{\bf{q}})
[d_{I}({\bf{q}}),a_{ {\bf{k}}\uparrow }(-{\bf{q}}\uparrow)]
 + \sum_{ {\bf{k}} }\Lambda_{e}({\bf{k}},-{\bf{q}})
[d_{I}({\bf{q}}),a^{\dagger}_{ {\bf{k}}\uparrow }({\bf{q}}\uparrow)]
\end{equation}
\begin{equation}
\rho^{(h)}({\bf{q}},I) = -\sum_{ {\bf{k}} }\Lambda_{h}({\bf{k}},{\bf{q}})
[d_{I}({\bf{q}}),a_{ {\bf{k}}\downarrow }(-{\bf{q}}\downarrow)]
 - \sum_{ {\bf{k}} }\Lambda_{h}({\bf{k}},-{\bf{q}})
[d_{I}({\bf{q}}),a^{\dagger}_{ {\bf{k}}\downarrow }({\bf{q}}\downarrow)]
\end{equation}
\begin{equation}
(\omega_{I\sigma}({\bf{q}}) - \Omega_{LO})
[d_{I\sigma}({\bf{q}}),b^{\dagger}_{ {\bf{q}} }]
= S(\sigma)\frac{ M_{ {\bf{q}} } }{\sqrt{V}}
\rho^{(\sigma)}({\bf{q}},I)
\end{equation}                
\begin{equation}
(\omega_{I\sigma}({\bf{q}}) + \Omega_{LO})[d_{I\sigma}({\bf{q}})
,b_{ -{\bf{q}} }]
= -S(\sigma)\frac{ M_{ {\bf{q}} } }{\sqrt{V}}
\rho^{(\sigma)}({\bf{q}},I)
\end{equation}
 From this we have the following fact that $ \omega_{I} $ are zeros of the
dielectric function,
\begin{equation}
\epsilon^{(e)}({\bf{q}},\omega^{(e)}_{I}) = 0 
\end{equation}
\begin{equation}
\epsilon^{(h)}({\bf{q}},\omega^{(h)}_{I}) = 0
\end{equation}
\begin{equation}
\epsilon^{(e,h)}({\bf{q}},\omega) = 1 - v({\bf{q}},\omega)
P^{(e,h)}({\bf{q}},\omega)                 
\end{equation}
and,
\begin{equation}
 v({\bf{q}},\omega) = v({\bf{q}}) +
 \frac{ 2 \Omega_{LO} M^{2}_{ {\bf{q}} } }{\omega^{2} - \Omega^{2}_{LO} }
\end{equation}
 and $ P^{(e,h)}({\bf{q}},\omega) $ is the usual RPA-polarization bubble.
 If we now make use of the fact that 
 {\mbox{ $  [d_{I\sigma}({\bf{q}}),d^{\dagger}_{I\sigma}({\bf{q}})] = 1 $ }}
 then we have,
\[
\sum_{ {\bf{k}} }|[d_{I\sigma}({\bf{q}}),
a^{\dagger}_{ {\bf{k}}\sigma}({\bf{q}} \sigma)]|^{2}
 - \sum_{ {\bf{k}} }|[d_{I\sigma}({\bf{q}}),
a_{ {\bf{k}}\sigma}(-{\bf{q}} \sigma)]|^{2}
\]
\begin{equation}
 + |[d_{I\sigma}({\bf{q}}),b^{\dagger}_{ {\bf{q}} }]|^{2}
 - |[d_{I\sigma}({\bf{q}}),b_{ -{\bf{q}} }]|^{2}
 = 1
\end{equation}
\[
\sum_{ {\bf{k}} }
\frac{ \Lambda^{2}_{\sigma}({\bf{k}},-{\bf{q}})
 - \Lambda^{2}_{\sigma}({\bf{k}},{\bf{q}})  }
{(\omega_{I\sigma}({\bf{q}}) -S(\sigma)\frac{ {\bf{k.q}} }
{m_{\sigma}})^{2}}[\rho^{(\sigma)}({\bf{q}},I)]^{2}
(\frac{ v({\bf{q}},I\sigma) }{V})^{2}
\]
\begin{equation}
 + \frac{ M^{2}_{ {\bf{q}} } }{V}
[\rho^{(\sigma)}({\bf{q}},I)]^{2}
\frac{ 4 \omega_{I\sigma}({\bf{q}})\Omega_{LO} }
{ (\omega^{2}_{I\sigma}({\bf{q}}) - \Omega^{2}_{LO})^{2} }
 = 1
\end{equation}
Therefore,
\begin{equation}
\rho^{(e)}({\bf{q}},I) = \{ (\frac{ v({\bf{q}},I,e) }{V})^{2}
\sum_{ {\bf{k}} }
\frac{ {\bar{n}}^{(e)}({\bf{k}}-{\bf{q}}/2)
 - {\bar{n}}^{(e)}({\bf{k}}+{\bf{q}}/2) }
{(\omega_{I,e}({\bf{q}}) -\frac{ {\bf{k.q}} }{m_{e}})^{2}}
 + \frac{ M^{2}_{ {\bf{q}} }  }{V}
\frac{ 4\omega_{I,e}({\bf{q}})\Omega_{LO} }
{(\omega^{2}_{I,e}({\bf{q}}) - \Omega^{2}_{LO})^{2} }\}^{-\frac{1}{2}}
\end{equation}
\begin{equation}
\rho^{(h)}({\bf{q}},I) = \{ (\frac{ v({\bf{q}},I,h) }{V})^{2}
\sum_{ {\bf{k}} }
\frac{ {\bar{n}}^{(h)}({\bf{k}}+{\bf{q}}/2)
 - {\bar{n}}^{(h)}({\bf{k}}-{\bf{q}}/2) }
{(\omega_{I,h}({\bf{q}}) +\frac{ {\bf{k.q}} }{m_{h}})^{2}}
 + \frac{ M^{2}_{ {\bf{q}} }  }{V}
\frac{ 4\omega_{I,h}({\bf{q}})\Omega_{LO} }
{ (\omega^{2}_{I,h}({\bf{q}}) - \Omega^{2}_{LO})^{2} } \}^{-\frac{1}{2}}
\end{equation}
Therefore,
\begin{equation}
\langle a^{\dagger}_{ {\bf{k}}\sigma }({\bf{q}}\sigma)
a_{ {\bf{k}}\sigma }({\bf{q}}\sigma) \rangle
 = \sum_{I} ([a_{ {\bf{k}}\sigma }({\bf{q}}\sigma),
d_{I\sigma}(-{\bf{q}})])^{2}
 = \sum_{I} (\frac{ v({\bf{q}},I\sigma) }{V})^{2}
\Lambda^{2}_{\sigma}({\bf{k}},-{\bf{q}})
\frac{ \rho^{2\sigma}(-{\bf{q}},I) }{(\omega_{I\sigma}({\bf{q}})
 + S(\sigma)\frac{ {\bf{k.q}} }{m_{\sigma}})^{2}}
\end{equation}
\begin{equation}
\langle a^{\dagger}_{ {\bf{k}}\uparrow }({\bf{q}}\uparrow)
a_{ {\bf{k}}\uparrow }({\bf{q}}\uparrow) \rangle
 = \sum_{I}(\frac{ v({\bf{q}},I,e) }{V})^{2}
\Lambda^{2}_{e}({\bf{k}},-{\bf{q}})
\frac{ (\rho^{(e)}(-{\bf{q}},I))^{2} }
{(\omega_{I,e}({\bf{q}})
 + \frac{ {\bf{k.q}} }{m_{e}})^{2}}
\label{ELECEQN2}
\end{equation}
\begin{equation}
\langle a^{\dagger}_{ {\bf{k}}\downarrow }({\bf{q}}\downarrow)
a_{ {\bf{k}}\downarrow }({\bf{q}}\downarrow) \rangle
 = \sum_{I}(\frac{ v({\bf{q}},I,h) }{V})^{2}
\Lambda^{2}_{h}({\bf{k}},-{\bf{q}})
\frac{ (\rho^{(h)}(-{\bf{q}},I))^{2} }
{(\omega_{I,h}({\bf{q}})
 - \frac{ {\bf{k.q}} }{m_{h}})^{2}}
\label{HOLEEQN2}
\end{equation}
 After some algebra it is clear that Eqs.(~\ref{ELECEQN1}) and
 (~\ref{HOLEEQN1}) are identical to Eqs.(~\ref{ELECEQN2})
 and (~\ref{HOLEEQN2}) respectively. Let us now solve the
 fundamental 
 equations namely Eq.(~\ref{FUNDEQ1}) and Eq.(~\ref{FUNDEQ2}). 
 For this we first would like to decompose the various fields in the exciton
 basis.
\begin{equation}
a_{ {\bf{k}}\downarrow }({\bf{0}}\uparrow)
 = \sum_{I} {\tilde{\varphi}}_{I}({\bf{k}})e^{-i\epsilon_{I} \mbox{ }t}
 {\tilde{D}}_{I}
\end{equation}
\begin{equation}
a_{ {\bf{k}}\uparrow }({\bf{0}}\downarrow)
 = e^{i(\epsilon^{e}({\bf{k}}) + \epsilon^{h}({\bf{k}}) ) \mbox{ }t}
 {\tilde{a}}_{ {\bf{k}}\uparrow }({\bf{0}}\downarrow)
\end{equation}
This means we may rewrite these equations as follows :
\[
i\frac{ \partial }{ \partial t }{\tilde{D}}_{I}(t)
 = 
 (\frac{|e|}{\mu c})e^{i\epsilon_{I} \mbox{ }t}
{\vec{A}}^{*}_{ext}(t).{\vec{p}}_{vc}
\sum_{ {\bf{k}} }
\Lambda_{1}({\bf{k}},{\bf{0}})
{\tilde{\varphi}}^{*}_{I}({\bf{k}})
\]
\[
 + \sum_{ {\bf{Q}} \neq 0 }\frac{ v_{eh}({\bf{Q}}) }{V}
[\sum_{ {\bf{k}}, J }{\tilde{\varphi}}^{*}_{I}({\bf{k}})
(1 - \Lambda_{1}({\bf{k}},{\bf{0}})\Lambda_{1}({\bf{k}}-{\bf{Q}},{\bf{0}}))
{\tilde{\varphi}}_{J}({\bf{k}}-{\bf{Q}})
e^{i(\epsilon_{I} - \epsilon_{J}) \mbox{   }t}
{\tilde{D}}_{J}(t)
\]
\begin{equation}
 - \sum_{ {\bf{k}} }{\tilde{\varphi}}^{*}_{I}({\bf{k}})
\Lambda_{1}({\bf{k}},{\bf{0}})\Lambda_{2}({\bf{k}}-{\bf{Q}},{\bf{0}})
e^{i\epsilon_{I} \mbox{   }t}
e^{-i(\epsilon^{e}({\bf{k}}-{\bf{Q}}) + \epsilon^{h}({\bf{k}}-{\bf{Q}}) )t}
{\tilde{a}}^{\dagger}_{ {\bf{k}} - {\bf{Q}} \uparrow }({\bf{0}}\downarrow)]
\label{EXBL1}
\end{equation}
\[
i\frac{ \partial }{ \partial t}
 {\tilde{a}}_{ {\bf{k}} \uparrow }({\bf{0}}\downarrow)
 = -\sum_{ {\bf{Q}} \neq 0 }\frac{ v_{eh}({\bf{Q}}) }{V}
\Lambda_{2}({\bf{k}},{\bf{0}})
[\Lambda_{1}({\bf{k}} + {\bf{Q}},{\bf{0}})
{\tilde{\varphi}}^{*}_{J}({\bf{k}}+{\bf{Q}}) 
e^{i\epsilon_{J} \mbox{      }t}
e^{-i(\epsilon^{e}({\bf{k}}) + \epsilon^{h}({\bf{k}})) \mbox{      }t}
{\tilde{D}}^{\dagger}_{J}
\]
\begin{equation}
+ \Lambda_{2}({\bf{k}} + {\bf{Q}},{\bf{0}})
e^{-i(\epsilon^{e}({\bf{k}}) + \epsilon^{h}({\bf{k}})) \mbox{      }t}
e^{i(\epsilon^{e}({\bf{k}}+{\bf{Q}})
 + \epsilon^{h}({\bf{k}}+{\bf{Q}})) \mbox{      }t}
{\tilde{a}}_{ {\bf{k}} + {\bf{Q}} \uparrow }({\bf{0}}\downarrow) ]
  + (\frac{|e|}{\mu c}){\vec{A}}_{ext}(t).{\vec{p}}_{vc}
\mbox{       }\Lambda_{2}({\bf{k}},{\bf{0}})
e^{-i(\epsilon^{e}({\bf{k}}) + \epsilon^{h}({\bf{k}})) \mbox{      }t}
\label{EXBL2}
\end{equation}


\[
i\frac{ \partial }{ \partial t }{\tilde{D}}_{0}(t)
 = 
 (\frac{|e|}{\mu c})e^{i\epsilon_{0} \mbox{ }t}
{\vec{A}}^{*}_{ext}(t).{\vec{p}}_{vc}
\sum_{ {\bf{k}} }
\Lambda_{1}({\bf{k}},{\bf{0}})
{\tilde{\varphi}}^{*}_{0}({\bf{k}})
\]
\[
 + \sum_{ {\bf{k}}^{'} \neq {\bf{k}} }
\frac{ v_{eh}({\bf{k}}-{\bf{k}}^{'}) }{V}
{\tilde{\varphi}}^{*}_{0}({\bf{k}})
(1 - \Lambda_{1}({\bf{k}},{\bf{0}})\Lambda_{1}({\bf{k}}^{'},{\bf{0}}))
{\tilde{\varphi}}_{0}({\bf{k}}^{'})
{\tilde{D}}_{0}(t)
\]
\[
 + \sum_{ {\bf{k}}^{'} \neq {\bf{k}} }
\frac{ v_{eh}({\bf{k}}-{\bf{k}}^{'}) }{V}
[{\tilde{\varphi}}^{*}_{0}({\bf{k}})
(1 - \Lambda_{1}({\bf{k}},{\bf{0}})\Lambda_{1}({\bf{k}}^{'},{\bf{0}}))
e^{i(\epsilon_{0} - \epsilon_{ {\bf{k}}^{'} }) \mbox{   }t}
{\tilde{D}}_{ {\bf{k}}^{'} }(t)
\]
\begin{equation}
 - {\tilde{\varphi}}^{*}_{0}({\bf{k}})
\Lambda_{1}({\bf{k}},{\bf{0}})\Lambda_{2}({\bf{k}}^{'},{\bf{0}})
e^{i(\epsilon_{0} - \epsilon_{ {\bf{k}}^{'} }) \mbox{   }t}
{\tilde{a}}^{\dagger}_{ {\bf{k}}^{'} \uparrow }({\bf{0}}\downarrow)]
\end{equation}
\[
i\frac{ \partial }{ \partial t }{\tilde{D}}_{ {\bf{k}} }(t)
 = 
 (\frac{|e|}{\mu c})e^{i\epsilon_{ {\bf{k}} } \mbox{ }t}
{\vec{A}}^{*}_{ext}(t).{\vec{p}}_{vc}
\Lambda_{1}({\bf{k}},{\bf{0}})
\]
\[
 + \sum_{ {\bf{k}}^{'} \neq {\bf{k}} }
\frac{ v_{eh}({\bf{k}}-{\bf{k}}^{'}) }{V}
(1 - \Lambda_{1}({\bf{k}},{\bf{0}})\Lambda_{1}({\bf{k}}^{'},{\bf{0}}))
{\tilde{\varphi}}_{0}({\bf{k}}^{'})
e^{i(\epsilon_{ {\bf{k}} } - \epsilon_{0}) \mbox{   }t}
{\tilde{D}}_{0}(t)
\]
\[
 + \sum_{ {\bf{k}}^{'} \neq {\bf{k}} }\frac{ v_{eh}({\bf{k}}-{\bf{k}}^{'}) }{V}
[(1 - \Lambda_{1}({\bf{k}},{\bf{0}})\Lambda_{1}({\bf{k}}^{'},{\bf{0}}))
e^{i(\epsilon_{ {\bf{k}} } - \epsilon_{ {\bf{k}}^{'} }) \mbox{   }t}
{\tilde{D}}_{ {\bf{k}}^{'} }(t)
\]
\begin{equation}
 - 
\Lambda_{1}({\bf{k}},{\bf{0}})\Lambda_{2}({\bf{k}}^{'},{\bf{0}})
e^{i(\epsilon_{ {\bf{k}} } - \epsilon_{ {\bf{k}}^{'} })\mbox{   }t}
{\tilde{a}}^{\dagger}_{ {\bf{k}}^{'} \uparrow }({\bf{0}}\downarrow)]
\end{equation}
\[
i\frac{ \partial }{ \partial t }
 {\tilde{a}}^{\dagger}_{ {\bf{k}} \uparrow }({\bf{0}}\downarrow)
 = \sum_{ {\bf{k}}^{'} \neq {\bf{k}} }\frac{ v_{eh}({\bf{k}}-{\bf{k}}^{'}) }{V}
\Lambda_{2}({\bf{k}},{\bf{0}})
\Lambda_{1}({\bf{k}}^{'},{\bf{0}})
{\tilde{\varphi}}_{0}({\bf{k}}^{'}) 
e^{-i(\epsilon_{0} - \epsilon_{ {\bf{k}} }) \mbox{      }t}
{\tilde{D}}_{0}(t)
\]
\[
+ \sum_{ {\bf{k}}^{'} \neq {\bf{k}} }\frac{ v_{eh}({\bf{k}}-{\bf{k}}^{'}) }{V}
\Lambda_{2}({\bf{k}},{\bf{0}})
[\Lambda_{1}({\bf{k}}^{'},{\bf{0}})
e^{-i(\epsilon_{ {\bf{k}}^{'} } - \epsilon_{ {\bf{k}} }) \mbox{      }t}
{\tilde{D}}_{ {\bf{k}}^{'} }(t)
\]
\begin{equation}
+ \Lambda_{2}({\bf{k}}^{'},{\bf{0}})
e^{i(\epsilon_{ {\bf{k}} } - \epsilon_{ {\bf{k}}^{'} })t}
{\tilde{a}}^{\dagger}_{ {\bf{k}}^{'} \uparrow }({\bf{0}}\downarrow) ]
  - (\frac{|e|}{\mu c}){\vec{A}}^{*}_{ext}(t).{\vec{p}}_{vc}
\mbox{       }\Lambda_{2}({\bf{k}},{\bf{0}})
e^{i \epsilon_{ {\bf{k}} } \mbox{      }t}
\end{equation}
\begin{equation}
{\tilde{\varphi}}_{0}({\bf{k}}) = \frac{1}{\sqrt{V}}
(4 \pi)\sqrt{ \frac{1}{\pi a^{3}_{X}} }
\frac{ 2\mbox{    }/a_{X} }{ (1/a^{2}_{X} + k^{2})^{2} }
\end{equation}
Furthermore,

\[
{\bar{n}}_{ e }({\bf{k}}) = {\bar{n}}_{ 0 }({\bf{k}})
+ \sum_{ I,{\bf{q}} }(\frac{ v({\bf{q}},I,e) }{V})^{2}
{\bar{n}}_{e}({\bf{k}}-{\bf{q}})(1 - {\bar{n}}_{e}({\bf{k}}))
\frac{ (\rho^{(e)}(-{\bf{q}},I))^{2} }
{(\omega_{I,e}({\bf{q}})
 + \frac{ {\bf{k.q}} }{m_{e}} - \frac{ q^{2} }{2m_{e}})^{2}}
\]
\begin{equation}
- \sum_{ I,{\bf{q}} }(\frac{ v({\bf{q}},I,e) }{V})^{2}
{\bar{n}}_{e}({\bf{k}})(1 - {\bar{n}}_{e}({\bf{k}}+{\bf{q}}))
\frac{ (\rho^{(e)}(-{\bf{q}},I))^{2} }
{(\omega_{I,e}({\bf{q}})
 + \frac{ {\bf{k.q}} }{m_{e}} + \frac{ q^{2} }{2m_{e}})^{2}}
\end{equation}

\[
{\bar{n}}_{ h }({\bf{k}}) = {\bar{n}}_{ 0 }({\bf{k}})
- \sum_{ I,{\bf{q}} }(\frac{ v({\bf{q}},I,h) }{V})^{2}
{\bar{n}}_{h}({\bf{k}})(1 - {\bar{n}}_{h}({\bf{k}}-{\bf{q}}))
\frac{ (\rho^{(h)}(-{\bf{q}},I))^{2} }
{(\omega_{I,h}({\bf{q}})
 - \frac{ {\bf{k.q}} }{m_{h}} + \frac{ q^{2} }{2m_{h}})^{2}}
\]
\begin{equation}
+ \sum_{ I,{\bf{q}} }(\frac{ v({\bf{q}},I,h) }{V})^{2}
{\bar{n}}_{h}({\bf{k}}+{\bf{q}})(1 - {\bar{n}}_{h}({\bf{k}}))
\frac{ (\rho^{(h)}(-{\bf{q}},I))^{2} }
{(\omega_{I,h}({\bf{q}})
 - \frac{ {\bf{k.q}} }{m_{h}} - \frac{ q^{2} }{2m_{h}})^{2}}
\end{equation}
\[
{\bar{n}}_{ 0 }({\bf{k}}) = |\varphi_{0}({\bf{k}})|^{2}
\langle {\tilde{D}}^{\dagger}_{0}(t){\tilde{D}}_{0}(t) \rangle
 + \langle {\tilde{D}}^{\dagger}_{ {\bf{k}} }(t)
{\tilde{D}}_{ {\bf{k}} }(t) \rangle
\]
\begin{equation}
+ \varphi_{0}({\bf{k}})e^{i(\epsilon_{ {\bf{k}} } - \epsilon_{0})t}
\langle {\tilde{D}}^{\dagger}_{ {\bf{k}} }(t)
{\tilde{D}}_{0}(t) \rangle
 + \varphi_{0}({\bf{k}})e^{-i(\epsilon_{ {\bf{k}} } - \epsilon_{0})t}
\langle {\tilde{D}}^{\dagger}_{0}(t){\tilde{D}}_{ {\bf{k}} }(t) \rangle
 - \langle {\tilde{a}}^{\dagger}_{ {\bf{k}} \uparrow }({\bf{0}}\downarrow)
{\tilde{a}}_{ {\bf{k}} \uparrow }({\bf{0}}\downarrow) \rangle
\end{equation}
or,
\[
{\bar{n}}_{0}({\bf{k}}) = |\varphi_{0}({\bf{k}})|^{2}
(\langle {\tilde{D}}^{r}_{0}(t){\tilde{D}}^{r}_{0}(t) \rangle
 + \langle {\tilde{D}}^{i}_{0}(t){\tilde{D}}^{i}_{0}(t) \rangle)
 + \langle {\tilde{D}}^{r}_{ {\bf{k}} }(t)
{\tilde{D}}^{r}_{ {\bf{k}} }(t) \rangle
 + \langle {\tilde{D}}^{i}_{ {\bf{k}} }(t)
{\tilde{D}}^{i}_{ {\bf{k}} }(t) \rangle
\]
\[
+ 2\varphi_{0}({\bf{k}})cos((\epsilon_{ {\bf{k}} } - \epsilon_{0})t)
[ \langle {\tilde{D}}^{r}_{ {\bf{k}} }(t)
{\tilde{D}}^{r}_{0}(t) \rangle
 + \langle {\tilde{D}}^{i}_{ {\bf{k}} }(t)
{\tilde{D}}^{i}_{0}(t) \rangle ]
\]
\begin{equation}
- 2\varphi_{0}({\bf{k}})sin((\epsilon_{ {\bf{k}} } - \epsilon_{0})t)
[\langle {\tilde{D}}^{r}_{ {\bf{k}} }(t)
{\tilde{D}}^{i}_{0}(t) \rangle
 -\langle {\tilde{D}}^{i}_{ {\bf{k}} }(t)
{\tilde{D}}^{r}_{0}(t) \rangle ]
 - \langle {\tilde{a}}^{r}_{ {\bf{k}}\uparrow  }({\bf{0}}\downarrow)
{\tilde{a}}^{r}_{ {\bf{k}}\uparrow  }({\bf{0}}\downarrow) \rangle
 -  \langle {\tilde{a}}^{i}_{ {\bf{k}}\uparrow  }({\bf{0}}\downarrow)
{\tilde{a}}^{i}_{ {\bf{k}}\uparrow  }({\bf{0}}\downarrow) \rangle
\label{NOEQN}
\end{equation}
\begin{equation}
{\bar{n}}_{e}({\bf{k}}) = {\bar{n}}_{0}({\bf{k}}) A_{e}({\bf{k}})
 + (1-{\bar{n}}_{0}({\bf{k}}))
B_{e}({\bf{k}}) 
\end{equation} 
\begin{equation}
{\bar{n}}_{h}({\bf{k}}) = {\bar{n}}_{0}({\bf{k}}) A_{h}({\bf{k}})
 + (1-{\bar{n}}_{0}({\bf{k}}))
B_{h}({\bf{k}}) 
\end{equation} 
\begin{equation}
 A_{e}({\bf{k}})
 = \frac{1}{ 1 + \frac{ T^{e}_{2}({\bf{k}}) }{ 1 +  T^{e}_{1}({\bf{k}}) } }
\end{equation}
\begin{equation}
 B_{e}({\bf{k}})
 = \frac{1}{ 1 + \frac{ 1 + T^{e}_{2}({\bf{k}}) }{ T^{e}_{1}({\bf{k}}) } }
\end{equation}
\begin{equation}
 T^{e}_{1}({\bf{k}}) =  \sum_{ I,{\bf{q}} }(\frac{ v({\bf{q}},I,e) }{V})^{2}
{\bar{n}}_{e}({\bf{k}}-{\bf{q}})
\frac{ (\rho^{(e)}(-{\bf{q}},I))^{2} }
{(\omega_{I,e}({\bf{q}})
 + \frac{ {\bf{k.q}} }{m_{e}} - \frac{ q^{2} }{2m_{e}})^{2} }
\end{equation}
\begin{equation}
 T^{e}_{2}({\bf{k}}) =  \sum_{I,{\bf{q}} }
(\frac{ v({\bf{q}},I,e) }{V})^{2}
(1 - {\bar{n}}_{e}({\bf{k}}+{\bf{q}}))
\frac{ (\rho^{(e)}(-{\bf{q}},I))^{2} }
{ (\omega_{I,e}({\bf{q}})
 + \frac{ {\bf{k.q}} }{m_{e}} + \frac{ q^{2} }{2m_{e}})^{2} }
\end{equation}
\begin{equation}
 A_{h}({\bf{k}})
 = \frac{1}{ 1 + \frac{ T^{h}_{1}({\bf{k}}) }{ 1 +  T^{h}_{2}({\bf{k}}) } }
\end{equation}
\begin{equation}
 B_{h}({\bf{k}})
 = \frac{1}{ 1 + \frac{ 1 + T^{h}_{1}({\bf{k}}) }{ T^{h}_{2}({\bf{k}}) } }
\end{equation}
\begin{equation}
 T^{h}_{1}({\bf{k}}) = 
 \sum_{ I,{\bf{q}} }(\frac{ v({\bf{q}},I,h) }{V})^{2}
(1-{\bar{n}}_{h}({\bf{k}}-{\bf{q}}))
\frac{ (\rho^{(h)}(-{\bf{q}},I))^{2} }
{(\omega_{I,h}({\bf{q}})
 - \frac{ {\bf{k.q}} }{m_{h}} + \frac{ q^{2} }{2m_{h}})^{2} }
\end{equation}
\begin{equation}
 T^{h}_{2}({\bf{k}}) =  \sum_{I,{\bf{q}} }
(\frac{ v({\bf{q}},I,h) }{V})^{2}
{\bar{n}}_{h}({\bf{k}}+{\bf{q}})
\frac{ (\rho^{(h)}(-{\bf{q}},I))^{2} }
{ (\omega_{I,h}({\bf{q}})
 - \frac{ {\bf{k.q}} }{m_{h}} - \frac{ q^{2} }{2m_{h}})^{2} }
\end{equation}
\[
\frac{ \partial }{ \partial t }{\tilde{D}}^{r}_{0}(t)
 = 
 -(\frac{|e|}{\mu c})
A_{X}^{i}(\epsilon_{0},t)p_{vc}
\sum_{ {\bf{k}} }
\Lambda_{1}({\bf{k}},{\bf{0}})
{\tilde{\varphi}}_{0}({\bf{k}})
\]
\[
 + \sum_{ {\bf{k}}^{'} \neq {\bf{k}} }
\frac{ v_{eh}({\bf{k}}-{\bf{k}}^{'}) }{V}
{\tilde{\varphi}}_{0}({\bf{k}})
(1 - \Lambda_{1}({\bf{k}},{\bf{0}})\Lambda_{1}({\bf{k}}^{'},{\bf{0}}))
{\tilde{\varphi}}_{0}({\bf{k}}^{'})
{\tilde{D}}^{i}_{0}(t)
\]
\[
 + \sum_{ {\bf{k}}^{'} \neq {\bf{k}} }
\frac{ v_{eh}({\bf{k}}-{\bf{k}}^{'}) }{V}
sin((\epsilon_{0} - \epsilon_{ {\bf{k}}^{'} }) \mbox{   }t)
[{\tilde{\varphi}}_{0}({\bf{k}})
(1 - \Lambda_{1}({\bf{k}},{\bf{0}})\Lambda_{1}({\bf{k}}^{'},{\bf{0}}))
{\tilde{D}}^{r}_{ {\bf{k}}^{'} }(t)
\]
\[
 - {\tilde{\varphi}}_{0}({\bf{k}})
\Lambda_{1}({\bf{k}},{\bf{0}})\Lambda_{2}({\bf{k}}^{'},{\bf{0}})
{\tilde{a}}^{r}_{ {\bf{k}}^{'} \uparrow }({\bf{0}}\downarrow)]
\]
\[
 + \sum_{ {\bf{k}}^{'} \neq {\bf{k}} }
\frac{ v_{eh}({\bf{k}}-{\bf{k}}^{'}) }{V}
cos((\epsilon_{0} - \epsilon_{ {\bf{k}}^{'} }) \mbox{   }t)
[{\tilde{\varphi}}_{0}({\bf{k}})
(1 - \Lambda_{1}({\bf{k}},{\bf{0}})\Lambda_{1}({\bf{k}}^{'},{\bf{0}}))
{\tilde{D}}^{i}_{ {\bf{k}}^{'} }(t)
\]
\begin{equation}
 + {\tilde{\varphi}}_{0}({\bf{k}})
\Lambda_{1}({\bf{k}},{\bf{0}})\Lambda_{2}({\bf{k}}^{'},{\bf{0}})
{\tilde{a}}^{i}_{ {\bf{k}}^{'} \uparrow }({\bf{0}}\downarrow)]
\end{equation}
\[
\frac{ \partial }{ \partial t }{\tilde{D}}^{i}_{0}(t)
 = 
 -(\frac{|e|}{\mu c})
A_{X}^{r}(\epsilon_{0},t)p_{vc}
\sum_{ {\bf{k}} }
\Lambda_{1}({\bf{k}},{\bf{0}})
{\tilde{\varphi}}_{0}({\bf{k}})
\]
\[
 - \sum_{ {\bf{k}}^{'} \neq {\bf{k}} }
\frac{ v_{eh}({\bf{k}}-{\bf{k}}^{'}) }{V}
{\tilde{\varphi}}_{0}({\bf{k}})
(1 - \Lambda_{1}({\bf{k}},{\bf{0}})\Lambda_{1}({\bf{k}}^{'},{\bf{0}}))
{\tilde{\varphi}}_{0}({\bf{k}}^{'})
{\tilde{D}}^{r}_{0}(t)
\]
\[
 - \sum_{ {\bf{k}}^{'} \neq {\bf{k}} }
\frac{ v_{eh}({\bf{k}}-{\bf{k}}^{'}) }{V}
cos((\epsilon_{0} - \epsilon_{ {\bf{k}}^{'} }) \mbox{   }t)
[{\tilde{\varphi}}_{0}({\bf{k}})
(1 - \Lambda_{1}({\bf{k}},{\bf{0}})\Lambda_{1}({\bf{k}}^{'},{\bf{0}}))
{\tilde{D}}^{r}_{ {\bf{k}}^{'} }(t)
\]
\[
 - {\tilde{\varphi}}_{0}({\bf{k}})
\Lambda_{1}({\bf{k}},{\bf{0}})\Lambda_{2}({\bf{k}}^{'},{\bf{0}})
{\tilde{a}}^{r}_{ {\bf{k}}^{'} \uparrow }({\bf{0}}\downarrow)]
\]
\[
 + \sum_{ {\bf{k}}^{'} \neq {\bf{k}} }
\frac{ v_{eh}({\bf{k}}-{\bf{k}}^{'}) }{V}
sin((\epsilon_{0} - \epsilon_{ {\bf{k}}^{'} }) \mbox{   }t)
[{\tilde{\varphi}}_{0}({\bf{k}})
(1 - \Lambda_{1}({\bf{k}},{\bf{0}})\Lambda_{1}({\bf{k}}^{'},{\bf{0}}))
{\tilde{D}}^{i}_{ {\bf{k}}^{'} }(t)
\]
\begin{equation}
 + {\tilde{\varphi}}_{0}({\bf{k}})
\Lambda_{1}({\bf{k}},{\bf{0}})\Lambda_{2}({\bf{k}}^{'},{\bf{0}})
{\tilde{a}}^{i}_{ {\bf{k}}^{'} \uparrow }({\bf{0}}\downarrow)]
\end{equation}
\[
\frac{ \partial }{ \partial t }{\tilde{D}}^{r}_{ {\bf{k}} }(t)
 = 
 -(\frac{|e|}{\mu c})
A_{X}^{i}(\epsilon_{ {\bf{k}} },t)p_{vc}
\Lambda_{1}({\bf{k}},{\bf{0}})
\]
\[
 + \sum_{ {\bf{k}}^{'} \neq {\bf{k}} }
\frac{ v_{eh}({\bf{k}}-{\bf{k}}^{'}) }{V}
(1 - \Lambda_{1}({\bf{k}},{\bf{0}})\Lambda_{1}({\bf{k}}^{'},{\bf{0}}))
{\tilde{\varphi}}_{0}({\bf{k}}^{'})
[sin((\epsilon_{ {\bf{k}} } - \epsilon_{0}) \mbox{   }t)
{\tilde{D}}^{r}_{0}(t)
+ cos((\epsilon_{ {\bf{k}} } - \epsilon_{0}) \mbox{   }t)
{\tilde{D}}^{i}_{0}(t)]
\]
\[
 + \sum_{ {\bf{k}}^{'} \neq {\bf{k}} }\frac{ v_{eh}({\bf{k}}-{\bf{k}}^{'}) }{V}
[(1 - \Lambda_{1}({\bf{k}},{\bf{0}})\Lambda_{1}({\bf{k}}^{'},{\bf{0}}))
\{sin((\epsilon_{ {\bf{k}} } - \epsilon_{ {\bf{k}}^{'} }) \mbox{   }t)
{\tilde{D}}^{r}_{ {\bf{k}}^{'} }(t)
+ cos((\epsilon_{ {\bf{k}} } - \epsilon_{ {\bf{k}}^{'} }) \mbox{   }t)
{\tilde{D}}^{i}_{ {\bf{k}}^{'} }(t)\}
\]
\begin{equation}
 - 
\Lambda_{1}({\bf{k}},{\bf{0}})\Lambda_{2}({\bf{k}}^{'},{\bf{0}})
\{sin((\epsilon_{ {\bf{k}} } - \epsilon_{ {\bf{k}}^{'} })\mbox{   }t)
{\tilde{a}}^{r}_{ {\bf{k}}^{'} \uparrow }({\bf{0}}\downarrow)
- cos((\epsilon_{ {\bf{k}} } - \epsilon_{ {\bf{k}}^{'} })\mbox{   }t)
{\tilde{a}}^{i}_{ {\bf{k}}^{'} \uparrow }({\bf{0}}\downarrow)
\}]
\end{equation}
\[
-\frac{ \partial }{ \partial t }{\tilde{D}}^{i}_{ {\bf{k}} }(t)
 = (\frac{|e|}{\mu c})
A^{r}_{X}(\epsilon_{ {\bf{k}} },t)p_{vc}
\Lambda_{1}({\bf{k}},{\bf{0}})
\]
\[
 + \sum_{ {\bf{k}}^{'} \neq {\bf{k}} }
\frac{ v_{eh}({\bf{k}}-{\bf{k}}^{'}) }{V}
(1 - \Lambda_{1}({\bf{k}},{\bf{0}})\Lambda_{1}({\bf{k}}^{'},{\bf{0}}))
{\tilde{\varphi}}_{0}({\bf{k}}^{'})
\{ cos((\epsilon_{ {\bf{k}} } - \epsilon_{0}) \mbox{   }t)
{\tilde{D}}^{r}_{0}(t)
 - sin((\epsilon_{ {\bf{k}} } - \epsilon_{0}) \mbox{   }t)
{\tilde{D}}^{i}_{0}(t) \}
\]
\[
 + \sum_{ {\bf{k}}^{'} \neq {\bf{k}} }\frac{ v_{eh}({\bf{k}}-{\bf{k}}^{'}) }{V}
[(1 - \Lambda_{1}({\bf{k}},{\bf{0}})\Lambda_{1}({\bf{k}}^{'},{\bf{0}}))
\{ cos((\epsilon_{ {\bf{k}} } - \epsilon_{ {\bf{k}}^{'} }) \mbox{   }t)
{\tilde{D}}^{r}_{ {\bf{k}}^{'} }(t)
 - sin((\epsilon_{ {\bf{k}} } - \epsilon_{ {\bf{k}}^{'} }) \mbox{   }t)
{\tilde{D}}^{i}_{ {\bf{k}}^{'} }(t) \}
\]
\begin{equation}
 - \Lambda_{1}({\bf{k}},{\bf{0}})\Lambda_{2}({\bf{k}}^{'},{\bf{0}})
\{cos((\epsilon_{ {\bf{k}} } - \epsilon_{ {\bf{k}}^{'} })\mbox{   }t)
{\tilde{a}}^{r}_{ {\bf{k}}^{'} \uparrow }({\bf{0}}\downarrow)
 + sin((\epsilon_{ {\bf{k}} } - \epsilon_{ {\bf{k}}^{'} })\mbox{   }t)
{\tilde{a}}^{i}_{ {\bf{k}}^{'} \uparrow }({\bf{0}}\downarrow)]
\end{equation}

\[
\frac{ \partial }{ \partial t }
 {\tilde{a}}^{i}_{ {\bf{k}} \uparrow }({\bf{0}}\downarrow)
 = \sum_{ {\bf{k}}^{'} \neq {\bf{k}} }
\frac{ v_{eh}({\bf{k}}-{\bf{k}}^{'}) }{V}
\Lambda_{2}({\bf{k}},{\bf{0}})
\Lambda_{1}({\bf{k}}^{'},{\bf{0}})
{\tilde{\varphi}}_{0}({\bf{k}}^{'}) 
\{ cos((\epsilon_{0} - \epsilon_{ {\bf{k}} }) \mbox{      }t)
{\tilde{D}}^{r}_{0}(t)
 + sin((\epsilon_{0} - \epsilon_{ {\bf{k}} }) \mbox{      }t)
{\tilde{D}}^{i}_{0}(t) \}
\]
\[
+ \sum_{ {\bf{k}}^{'} \neq {\bf{k}} }\frac{ v_{eh}({\bf{k}}-{\bf{k}}^{'}) }{V}
\Lambda_{2}({\bf{k}},{\bf{0}})
[\Lambda_{1}({\bf{k}}^{'},{\bf{0}})
\{ cos((\epsilon_{ {\bf{k}}^{'} } - \epsilon_{ {\bf{k}} }) \mbox{      }t)
{\tilde{D}}^{r}_{ {\bf{k}}^{'} }(t)
 + sin((\epsilon_{ {\bf{k}}^{'} } - \epsilon_{ {\bf{k}} }) \mbox{      }t)
{\tilde{D}}^{i}_{ {\bf{k}}^{'} }(t) \}
\]
\begin{equation}
+ \Lambda_{2}({\bf{k}}^{'},{\bf{0}})
\{ cos((\epsilon_{ {\bf{k}} } - \epsilon_{ {\bf{k}}^{'} })t)
{\tilde{a}}^{r}_{ {\bf{k}}^{'} \uparrow }({\bf{0}}\downarrow)
+ sin((\epsilon_{ {\bf{k}} } - \epsilon_{ {\bf{k}}^{'} })t)
{\tilde{a}}^{i}_{ {\bf{k}}^{'} \uparrow }({\bf{0}}\downarrow) ]
  - (\frac{|e|}{\mu c})A^{r}_{X}(\epsilon_{ {\bf{k}} },t)p_{vc}
\mbox{       }\Lambda_{2}({\bf{k}},{\bf{0}})
\end{equation}

\[
\frac{ \partial }{ \partial t }
 {\tilde{a}}^{r}_{ {\bf{k}} \uparrow }({\bf{0}}\downarrow)
 = \sum_{ {\bf{k}}^{'} \neq {\bf{k}} }
\frac{ v_{eh}({\bf{k}}-{\bf{k}}^{'}) }{V}
\Lambda_{2}({\bf{k}},{\bf{0}})
\Lambda_{1}({\bf{k}}^{'},{\bf{0}})
{\tilde{\varphi}}_{0}({\bf{k}}^{'}) 
\{ -sin((\epsilon_{0} - \epsilon_{ {\bf{k}} }) \mbox{      }t)
{\tilde{D}}^{r}_{0}(t)
+ cos((\epsilon_{0} - \epsilon_{ {\bf{k}} }) \mbox{      }t)
{\tilde{D}}^{i}_{0}(t) \}
\]
\[
+ \sum_{ {\bf{k}}^{'} \neq {\bf{k}} }
\frac{ v_{eh}({\bf{k}}-{\bf{k}}^{'}) }{V}
\Lambda_{2}({\bf{k}},{\bf{0}})
[\Lambda_{1}({\bf{k}}^{'},{\bf{0}})
\{ -sin((\epsilon_{ {\bf{k}}^{'} } - \epsilon_{ {\bf{k}} }) \mbox{      }t)
{\tilde{D}}^{r}_{ {\bf{k}}^{'} }(t)
 + cos((\epsilon_{ {\bf{k}}^{'} } - \epsilon_{ {\bf{k}} }) \mbox{      }t)
{\tilde{D}}^{i}_{ {\bf{k}}^{'} }(t) \}
\]
\[
+ \Lambda_{2}({\bf{k}}^{'},{\bf{0}})
\{ sin((\epsilon_{ {\bf{k}} } - \epsilon_{ {\bf{k}}^{'} })t)
{\tilde{a}}^{r}_{ {\bf{k}}^{'} \uparrow }({\bf{0}}\downarrow)
  - cos((\epsilon_{ {\bf{k}} } - \epsilon_{ {\bf{k}}^{'} })t)
{\tilde{a}}^{i}_{ {\bf{k}}^{'} \uparrow }({\bf{0}}\downarrow) \} ]
\]
\begin{equation}
  + (\frac{|e|}{\mu c})A^{i}_{X}(\epsilon_{ {\bf{k}} },t)p_{vc}
\mbox{       }\Lambda_{2}({\bf{k}},{\bf{0}})
\end{equation}
In order to simplify the calculations further, let us define,
\[
P_{i}({\bf{q}},\omega) = (\frac{1}{4\pi})\int^{\infty}_{0}
dk\mbox{          }k^{2}\mbox{     }{\bar{n}}({\bf{k}})
(-\frac{m}{|k||q|})
[\theta(\omega-\frac{|k||q|}{m} - \frac{q^{2}}{2m})
 - \theta(\omega+\frac{|k||q|}{m} - \frac{q^{2}}{2m})
\]
\begin{equation}
 - \theta(\omega-\frac{|k||q|}{m} + \frac{q^{2}}{2m})
 + \theta(\omega+\frac{|k||q|}{m} + \frac{q^{2}}{2m})]
\end{equation}
\begin{equation}
P_{r}({\bf{q}},\omega) = \int^{\infty}_{0}
\frac{ d\omega^{'} }{\pi}\mbox{           }
P_{i}({\bf{q}},\omega^{'})
(\frac{2\omega^{'} }{ \omega^{'2} - \omega^{2} })
\end{equation}
\begin{equation}
S({\bf{q}},\omega) \hspace{.2in}\tilde{ }\hspace{.2in}
 \frac{ P_{i}({\bf{q}},\omega) }
{ (1 - v({\bf{q}},\omega)P_{r}({\bf{q}},\omega))^{2} 
 + v^{2}({\bf{q}},\omega)(P_{i}({\bf{q}},\omega))^{2} }
\end{equation}
\begin{equation}
\rho({\bf{q}},\omega) = \{ -(\frac{ v({\bf{q}},\omega) }{V})^{2}
V\int^{\infty}_{0}\frac{ d\omega^{'} }{\pi}
P_{i}({\bf{q}},\omega^{'})\frac{ 4\omega \omega^{'} }
{ (\omega^{'2} - \omega^{2})^{2} }
 + \frac{ M^{2}_{ {\bf{q}} } }{V}
\frac{ 4\omega \Omega_{LO} }
{ (\omega^{2} - \Omega^{2}_{LO})^{2} } \}^{-\frac{1}{2}}
\end{equation}

\section{Comparison with Semiconductor Bloch Equations}

 The equations presented in the previous section namely Eqs.(~\ref{EXBL1})
 and (~\ref{EXBL2}) are intended as alternatives to the usual Semiconductor
 Bloch equations used to study semiconductors.  Let us now try to solve this
 system with a pulse field.
 That is, we apply an external field with central frequency
 $ \omega_{X} $ and assume it lasts for a time 
 $ \tau_{X} $ and therefore we have a spread in frequency
 $ \Gamma_{X} = 2\pi/\tau_{X} $.  We would like to
 see the evolution of the polarization and populations in this case.
 This exercise also enables us to compare our results with those of the
 SBEs and ascertain where the differences lie. 
 To this end let us set,
\begin{equation}
{\vec{A}}_{ext}(t) = {\hat{p}}_{vc}A_{X}(0,t)
\end{equation}
where,
\begin{equation}
A_{X}(E,t) = A_{0}\int_{-\infty}^{\infty} d\omega\mbox{       }
\frac{ \Gamma_{X}/\pi }{(\omega - \omega_{X})^{2} + \Gamma^{2}_{X} }
e^{i(\omega-E) \mbox{  }t}
\end{equation}
For comparison the SBE is reproduced below.
\[
g_{hh}({\bf{k}}t)=i{\bar{n}}_{h}({\bf{k}})
 = i\langle d^{\dagger}_{ -{\bf{k}} }d_{ -{\bf{k}} }\rangle,
 \mbox{         } 
 = i \mbox{         }f({\bf{k}})
 g_{ee}({\bf{k}}t) = i{\bar{n}}_{e}({\bf{k}})
 = i\langle c^{\dagger}_{ {\bf{k}} }c_{ {\bf{k}} }\rangle
\]
\[
g_{he}({\bf{k}}t)=i\langle d_{ -{\bf{k}} }(t)c_{ {\bf{k}} }(t) \rangle
 = i\mbox{        }p({\bf{k}})
\]
\[
i\frac{\partial}{\partial t}g_{hh}({\bf{k}}t)
 = 2\mbox{         }
Re(\Omega({\bf{k}}t)g^{*}_{he}({\bf{k}}t)) + R_{hh}({\bf{k}}t)
\]
\[
i\frac{\partial}{\partial t}g_{he}({\bf{k}}t)
= -\Omega({\bf{k}}t)(i-2g_{hh}({\bf{k}}t))
 + (\epsilon_{h}({\bf{k}})+\epsilon_{c}({\bf{k}})-2\Sigma({\bf{k}}t))
g_{he}({\bf{k}}t) + R_{he}({\bf{k}}t)
\]
\[
\Omega({\bf{k}}t) = {\frac{|e|}{ \mu c}}{\bf{A}}^{*}_{ext}(t).{\bf{p_{vc}}}
- i\sum_{ {\bf{k}}^{'} }v_{ {\bf{k}}-{\bf{k}}^{'} }g_{he}({\bf{k^{'}}}t)
\]
\begin{equation}
\Sigma({\bf{k}}t) = -i\sum_{ { \bf{k^{'}} } }v_{ {\bf{k}}-{\bf{k}}^{'} }
g_{hh}({\bf{k}}^{'}t)
\end{equation}
Define,
\begin{equation}
\Omega({\bf{k}},t) = {\tilde{\Omega}}({\bf{k}},t)
e^{-i(k^{2}/2\mu + E_{g})t},
\mbox{           }p({\bf{k}},t) = {\tilde{p}}({\bf{k}},t)
e^{-i(k^{2}/2\mu + E_{g})t}
\end{equation}
\begin{equation}
\frac{ \partial f({\bf{k}}) }{\partial t}
 = 2{\tilde{\Omega}}_{R}({\bf{k}},t){\tilde{p}}_{I}({\bf{k}})
 - 2{\tilde{\Omega}}_{I}({\bf{k}},t){\tilde{p}}_{R}({\bf{k}})
\end{equation}
\begin{equation}
\frac{ \partial {\tilde{p}}_{R}({\bf{k}}) }{\partial t}
 = {\tilde{\Omega}}_{R}({\bf{k}},t)(1-2 f({\bf{k}}))
 - 2 \Sigma({\bf{k}},t) {\tilde{p}}_{I}({\bf{k}},t)
\end{equation}
\begin{equation}
\frac{ \partial {\tilde{p}}_{I}({\bf{k}}) }{\partial t}
 = {\tilde{\Omega}}_{I}({\bf{k}},t)(1-2 f({\bf{k}}))
 + 2 \Sigma({\bf{k}},t) {\tilde{p}}_{R}({\bf{k}},t)
\end{equation}
\begin{equation}
{\tilde{\Omega}}_{R}({\bf{k}},t) = (\frac{|e|}{\mu c})
{\tilde{A}}_{ext}(t)p_{vc}\mbox{         }
cos((k^{2}/2 \mu + E_{g}-\omega_{X})t)
 + \sum_{ {\bf{k}}^{'} \neq {\bf{k}} }
v_{ {\bf{k}} - {\bf{k}}^{'} }
[{\tilde{p}}_{R}({\bf{k}}^{'})
cos((k^{2}-k^{'2})t/2\mu)
 - {\tilde{p}}_{I}({\bf{k}}^{'})
sin((k^{2}-k^{'2})t/2\mu)]
\end{equation}
\begin{equation}
{\tilde{\Omega}}_{I}({\bf{k}},t) = (\frac{|e|}{\mu c})
{\tilde{A}}_{ext}(t)p_{vc}\mbox{         }
sin((k^{2}/2 \mu + E_{g}-\omega_{X})t)
 + \sum_{ {\bf{k}}^{'} \neq {\bf{k}} }
v_{ {\bf{k}} - {\bf{k}}^{'} }
[{\tilde{p}}_{I}({\bf{k}}^{'})
cos((k^{2}-k^{'2})t/2\mu)
 + {\tilde{p}}_{R}({\bf{k}}^{'})
sin((k^{2}-k^{'2})t/2\mu)]
\end{equation}
\begin{equation}
\Sigma({\bf{k}},t) = \sum_{ {\bf{k}}^{'} \neq {\bf{k}} }
v_{ {\bf{k}} - {\bf{k}}^{'} }f({\bf{k}}^{'})
\end{equation}

\section{Optical Conductivity}

First define the total polarization,
\begin{equation}
P(t) = \sum_{ {\bf{k}} }\langle d_{ -{\bf{k}} }c_{ {\bf{k}} }\rangle
\end{equation}
From this we may obtain the Fourier component,
\begin{equation}
{\tilde{P}}(\omega) = \int^{\infty}_{-\infty} dt \mbox{         }
P(t)\mbox{        }e^{i\omega t}
\end{equation}
Now since,
\begin{equation}
A_{\tau}(t) = A_{\tau} \mbox{       }e^{-i\mbox{  }\omega_{X}\mbox{  }\tau}
\delta(t-\tau)
\end{equation}
\begin{equation}
E(t) = -\frac{ \partial A_{\tau}(t) }{\partial t},
\mbox{        }-A_{\tau}e^{-i\omega_{X} \mbox{  }\tau}
 \mbox{       }\delta^{'}(t-\tau)
 = \int^{\infty}_{-\infty} \frac{ d\omega }{2\pi} \mbox{         }
e^{-i\omega t}E(\omega)
\end{equation}
\begin{equation}
E(\omega) = (i\omega) A_{\tau}\mbox{        }
 e^{i(\omega - \omega_{X})\tau}
\end{equation}
Since,
\begin{equation}
j(\omega) = p_{vc} {\tilde{P}}(\omega) 
\end{equation}
We have 
\begin{equation}
\sigma(\omega) =p_{vc} e^{-i(\omega - \omega_{X})\tau}
Lim_{ A_{\tau} \rightarrow 0 }
\frac{ ({\tilde{P}}(\omega, A_{\tau})-{\tilde{P}}(\omega, 0)) }
{ (i\omega) A_{\tau} }
\end{equation}
\newpage
Here,
\[
{\tilde{P}}(\omega) = \sum_{ {\bf{k}} }
\int^{\tau}_{-\infty} dt \mbox{        }
\sqrt{ (1- {\bar{n}}_{e}({\bf{k}}) )(1 - {\bar{n}}_{h}({\bf{k}})) }
[\varphi_{0}({\bf{k}}){\tilde{D}}^{(-)}_{0}(t)
e^{i(\omega-\epsilon_{0})t}
 + {\tilde{D}}^{(-)}_{ {\bf{k}} }(t)
e^{i(\omega-\epsilon_{ {\bf{k}} })t}]
\]
\[
+ \sum_{ {\bf{k}} }\int^{\tau}_{-\infty} dt \mbox{        }
\sqrt{ {\bar{n}}_{e}({\bf{k}}){\bar{n}}_{h}({\bf{k}})  }
e^{i(\omega-\epsilon_{ {\bf{k}} })t}
{\tilde{a}}^{\dagger}_{ {\bf{k}} \uparrow }(0,\downarrow,-)
\]
\[
+ \sum_{ {\bf{k}} }
\int^{\infty}_{\tau} dt \mbox{        }
\sqrt{ (1- {\bar{n}}_{e}({\bf{k}}))(1 - {\bar{n}}_{h}({\bf{k}})) }
[\varphi_{0}({\bf{k}}){\tilde{D}}^{(+)}_{0}(t)
e^{i(\omega-\epsilon_{0})t}
 + {\tilde{D}}^{(+)}_{ {\bf{k}} }(t)
e^{i(\omega-\epsilon_{ {\bf{k}} })t }]
\]
\begin{equation}
+ \sum_{ {\bf{k}} }\int_{\tau}^{\infty} dt \mbox{        }
\sqrt{ {\bar{n}}_{e}({\bf{k}}){\bar{n}}_{h}({\bf{k}})  }
e^{i(\omega-\epsilon_{ {\bf{k}} })t}
{\tilde{a}}^{\dagger}_{ {\bf{k}} \uparrow }(0,\downarrow,+)
\end{equation}
\begin{equation}
i({\tilde{D}}^{(+)}_{0}(\tau) -  {\tilde{D}}^{(-)}_{0}(\tau))
 = (\frac{ |e| }{\mu c})e^{i (\epsilon_{0} - \omega_{X})\tau }
A_{\tau} p_{vc} 
\sum_{ {\bf{k}} }\varphi_{0}({\bf{k}}) \Lambda_{1}({\bf{k}},{\bf{0}})
\end{equation}
\begin{equation}
i({\tilde{D}}^{(+)}_{ {\bf{k}} }(\tau) -  {\tilde{D}}^{(-)}_{ {\bf{k}} }(\tau))
 = (\frac{ |e| }{\mu c})e^{i (\frac{ k^{2} }{2\mu} + E_{g} - \omega_{X}) \tau }
A_{\tau} p_{vc}  \Lambda_{1}({\bf{k}},{\bf{0}})
\end{equation}
\begin{equation}
i({\tilde{a}}_{ {\bf{k}}  \uparrow}({\bf{0}}\downarrow,+)
 -  {\tilde{a}}_{ {\bf{k}}  \uparrow }({\bf{0}}\downarrow,-))
 = (\frac{ |e| }{\mu c})e^{-i (\frac{ k^{2} }{2\mu} + E_{g} -\omega_{X}) \tau }
A_{\tau} p_{vc}  \Lambda_{2}({\bf{k}},{\bf{0}})
\end{equation}
 \[
 {\tilde{P}}_{R}(\omega) = \int_{0}^{\infty}\frac{ 4 \pi k^{2} }{(2\pi)^{3}}
 dk \mbox{          }
 \int^{\tau}_{-\infty} dt \mbox{        }
 \sqrt{ (1- {\bar{n}}_{e}({\bf{k}}) )(1 - {\bar{n}}_{h}({\bf{k}})) }
 \]
 \[
 [\varphi_{0}({\bf{k}}){\tilde{D}}^{(-),R}_{0}(t)
 cos((\omega-\epsilon_{0})t)
 - \varphi_{0}({\bf{k}}){\tilde{D}}^{(-),I}_{0}(t)
 sin((\omega-\epsilon_{0})t)
 + {\tilde{D}}^{(-),R}_{ {\bf{k}} }(t)
 cos((\omega-\epsilon_{ {\bf{k}} })t)
 - {\tilde{D}}^{(-),I}_{ {\bf{k}} }(t)
 sin((\omega-\epsilon_{ {\bf{k}} })t)]
 \]
 \[
 +  \int_{0}^{\infty}\frac{ 4 \pi k^{2} }{(2\pi)^{3}}
 dk \mbox{          }\int^{\tau}_{-\infty} dt \mbox{        }
 \sqrt{ {\bar{n}}_{e}({\bf{k}}){\bar{n}}_{h}({\bf{k}})  }
 cos((\omega-\epsilon_{ {\bf{k}} })t)
 {\tilde{a}}^{R}_{ {\bf{k}} \uparrow }(0,\downarrow,-)
 \]
 \[
  + \int_{0}^{\infty}\frac{ 4 \pi k^{2} }{(2\pi)^{3}}
 dk \mbox{          }\int^{\tau}_{-\infty} dt \mbox{        }
 \sqrt{ {\bar{n}}_{e}({\bf{k}}){\bar{n}}_{h}({\bf{k}})  }
 sin((\omega-\epsilon_{ {\bf{k}} })t)
 {\tilde{a}}^{I}_{ {\bf{k}} \uparrow }(0,\downarrow,-)
 \]
 \[
 +
 \int_{0}^{\infty}\frac{ 4 \pi k^{2} }{(2\pi)^{3}}
 dk \mbox{          }
 \int^{\infty}_{\tau} dt \mbox{        }
 \sqrt{ (1- {\bar{n}}_{e}({\bf{k}}) )(1 - {\bar{n}}_{h}({\bf{k}})) }
 \]
 \[
 [\varphi_{0}({\bf{k}}){\tilde{D}}^{(+),R}_{0}(t)
 cos((\omega-\epsilon_{0})t)
  - \varphi_{0}({\bf{k}}){\tilde{D}}^{(+),I}_{0}(t)
 sin((\omega-\epsilon_{0})t)
 + {\tilde{D}}^{(+),R}_{ {\bf{k}} }(t)
 cos((\omega-\epsilon_{ {\bf{k}} })t)
 - {\tilde{D}}^{(+),I}_{ {\bf{k}} }(t)
 sin((\omega-\epsilon_{ {\bf{k}} })t)]
 \]
 \[
 +  \int_{0}^{\infty}\frac{ 4 \pi k^{2} }{(2\pi)^{3}}
 dk \mbox{          }\int^{\infty}_{\tau} dt \mbox{        }
 \sqrt{ {\bar{n}}_{e}({\bf{k}}){\bar{n}}_{h}({\bf{k}})  }
 cos((\omega-\epsilon_{ {\bf{k}} })t)
 {\tilde{a}}^{R}_{ {\bf{k}} \uparrow }(0,\downarrow,+)
 \]
 \begin{equation}
  + \int_{0}^{\infty}\frac{ 4 \pi k^{2} }{(2\pi)^{3}}
 dk \mbox{          }\int^{\infty}_{\tau} dt \mbox{        }
 \sqrt{ {\bar{n}}_{e}({\bf{k}}){\bar{n}}_{h}({\bf{k}})  }
 sin((\omega-\epsilon_{ {\bf{k}} })t)
 {\tilde{a}}^{I}_{ {\bf{k}} \uparrow }(0,\downarrow,+)
 \end{equation}

 \[
 {\tilde{P}}_{I}(\omega) =  \int_{0}^{\infty}\frac{ 4 \pi k^{2} }{(2\pi)^{3}}
 dk \mbox{          }
 \int^{\tau}_{-\infty} dt \mbox{        }
 \sqrt{ (1- {\bar{n}}_{e}({\bf{k}}) )(1 - {\bar{n}}_{h}({\bf{k}})) }
 \]
 \[
 [\varphi_{0}({\bf{k}}){\tilde{D}}^{(-),R}_{0}(t)
 sin((\omega-\epsilon_{0})t)
 + \varphi_{0}({\bf{k}}){\tilde{D}}^{(-),I}_{0}(t)
 cos((\omega-\epsilon_{0})t)
 + {\tilde{D}}^{(-),R}_{ {\bf{k}} }(t)
 sin((\omega-\epsilon_{ {\bf{k}} })t)
 + {\tilde{D}}^{(-),I}_{ {\bf{k}} }(t)
 cos((\omega-\epsilon_{ {\bf{k}} })t)]
 \]
 \[
 + \int^{\tau}_{-\infty} dt \mbox{        }
 \int_{0}^{\infty}\frac{ 4 \pi k^{2} }{(2\pi)^{3}}
 dk \mbox{          }
 \sqrt{ {\bar{n}}_{e}({\bf{k}}){\bar{n}}_{h}({\bf{k}})  }
 sin((\omega-\epsilon_{ {\bf{k}} })t)
 {\tilde{a}}^{R}_{ {\bf{k}} \uparrow }(0,\downarrow,-)
 \]
 \[
 - \int_{0}^{\infty}\frac{ 4 \pi k^{2} }{(2\pi)^{3}}
 dk \mbox{          }\int^{\tau}_{-\infty} dt \mbox{        }
 \sqrt{ {\bar{n}}_{e}({\bf{k}}){\bar{n}}_{h}({\bf{k}})  }
 cos((\omega-\epsilon_{ {\bf{k}} })t)
 {\tilde{a}}^{I}_{ {\bf{k}} \uparrow }(0,\downarrow,-)
 \]
 \[
 +  \int_{0}^{\infty}\frac{ 4 \pi k^{2} }{(2\pi)^{3}}
 dk \mbox{          }
 \int^{\infty}_{\tau} dt \mbox{        }
 \sqrt{ (1- {\bar{n}}_{e}({\bf{k}}))(1 - {\bar{n}}_{h}({\bf{k}})) }
 \]
 \[
 [\varphi_{0}({\bf{k}}){\tilde{D}}^{(+),R}_{0}(t)
 sin((\omega-\epsilon_{0})t)
 + \varphi_{0}({\bf{k}}){\tilde{D}}^{(+),I}_{0}(t)
 cos((\omega-\epsilon_{0})t)
  + {\tilde{D}}^{(+),R}_{ {\bf{k}} }(t)
 sin((\omega-\epsilon_{ {\bf{k}} })t )
 + {\tilde{D}}^{(+),I}_{ {\bf{k}} }(t)
 cos((\omega-\epsilon_{ {\bf{k}} })t )]
 \]
 \[
 +  \int_{0}^{\infty}\frac{ 4 \pi k^{2} }{(2\pi)^{3}}
 dk \mbox{          }\int_{\tau}^{\infty} dt \mbox{        }
 \sqrt{ {\bar{n}}_{e}({\bf{k}}){\bar{n}}_{h}({\bf{k}})  }
 sin((\omega-\epsilon_{ {\bf{k}} })t)
 {\tilde{a}}^{R}_{ {\bf{k}} \uparrow }(0,\downarrow,+)
 \]
 \begin{equation}
 -  \int_{0}^{\infty}\frac{ 4 \pi k^{2} }{(2\pi)^{3}}
 dk \mbox{          }\int_{\tau}^{\infty} dt \mbox{        }
 \sqrt{ {\bar{n}}_{e}({\bf{k}}){\bar{n}}_{h}({\bf{k}})  }
 cos((\omega-\epsilon_{ {\bf{k}} })t)
 {\tilde{a}}^{I}_{ {\bf{k}} \uparrow }(0,\downarrow,+)
 \end{equation}
\begin{equation}
 {\tilde{D}}_{0}^{(+),R}(\tau) - {\tilde{D}}_{0}^{(-),R}(\tau)
 = (\frac{ |e| }{\mu c})A_{\tau} p_{vc}
 sin( (\epsilon_{0} - \omega_{X})\tau)
 \int_{0}^{\infty}\frac{ 4 \pi k^{2} }{(2\pi)^{3}}
 dk \mbox{          }
 \varphi_{0}({\bf{k}})\Lambda_{1}({\bf{k}},{\bf{0}})
 \end{equation}

\begin{equation}
{\tilde{D}}_{0}^{(+),I}(\tau) - {\tilde{D}}_{0}^{(-),I}(\tau)
 = -(\frac{ |e| }{\mu c})A_{\tau} p_{vc}
cos( (\epsilon_{0} - \omega_{X})\tau)
 \int_{0}^{\infty}\frac{ 4 \pi k^{2} }{(2\pi)^{3}}
dk \mbox{          }
\varphi_{0}({\bf{k}})\Lambda_{1}({\bf{k}},{\bf{0}})
\end{equation}

\begin{equation}
{\tilde{D}}_{ {\bf{k}} }^{(+),R}(\tau) - {\tilde{D}}_{ {\bf{k}} }^{(-),R}(\tau)
 = (\frac{ |e| }{\mu c})A_{\tau} p_{vc}
sin( (\epsilon_{ {\bf{k}} } - \omega_{X})\tau)
\Lambda_{1}({\bf{k}},{\bf{0}})
\end{equation}

\begin{equation}
{\tilde{D}}_{ {\bf{k}} }^{(+),I}(\tau) - {\tilde{D}}_{ {\bf{k}} }^{(-),I}(\tau)
 = -(\frac{ |e| }{\mu c})A_{\tau} p_{vc}
cos( (\epsilon_{ {\bf{k}} } - \omega_{X})\tau)
\Lambda_{1}({\bf{k}},{\bf{0}})
\end{equation}

\begin{equation}
{\tilde{a}}_{ {\bf{k}} \uparrow }^{R}({\bf{0}},\downarrow,+)
 - {\tilde{a}}_{ {\bf{k}} \uparrow }^{R}({\bf{0}},\downarrow,-)
 = -(\frac{ |e| }{\mu c})A_{\tau} p_{vc}
sin( (\epsilon_{ {\bf{k}} } - \omega_{X})\tau)
\Lambda_{2}({\bf{k}},{\bf{0}})
\end{equation}

\begin{equation}
{\tilde{a}}_{ {\bf{k}} \uparrow }^{I}({\bf{0}},\downarrow,+)
 - {\tilde{a}}_{ {\bf{k}} \uparrow }^{I}({\bf{0}},\downarrow,-)
 = -(\frac{ |e| }{\mu c})A_{\tau} p_{vc}
cos( (\epsilon_{ {\bf{k}} } - \omega_{X})\tau)
\Lambda_{2}({\bf{k}},{\bf{0}})
\end{equation}

\[
{\tilde{P}}(\omega) = \sum_{ {\bf{k}} }
\int^{\tau}_{-\infty} dt \mbox{        }
\sqrt{ (1- {\bar{n}}_{e}({\bf{k}}) )(1 - {\bar{n}}_{h}({\bf{k}})) }
[\varphi_{0}({\bf{k}}){\tilde{D}}^{(-)}_{0}(t)
e^{i(\omega-\epsilon_{0})t}
 + {\tilde{D}}^{(-)}_{ {\bf{k}} }(t)
e^{i(\omega-\epsilon_{ {\bf{k}} })t}]
\]
\[
+ \sum_{ {\bf{k}} }\int^{\tau}_{-\infty} dt \mbox{        }
\sqrt{ {\bar{n}}_{e}({\bf{k}}){\bar{n}}_{h}({\bf{k}})  }
e^{i(\omega-\epsilon_{ {\bf{k}} })t}
{\tilde{a}}^{\dagger}_{ {\bf{k}} \uparrow }(0,\downarrow,-)
\]
\[
+ \sum_{ {\bf{k}} }
\int^{\infty}_{\tau} dt \mbox{        }
\sqrt{ (1- {\bar{n}}_{e}({\bf{k}}))(1 - {\bar{n}}_{h}({\bf{k}})) }
[\varphi_{0}({\bf{k}}){\tilde{D}}^{(+)}_{0}(t)
e^{i(\omega-\epsilon_{0})t}
 + {\tilde{D}}^{(+)}_{ {\bf{k}} }(t)
e^{i(\omega-\epsilon_{ {\bf{k}} })t }]
\]
\begin{equation}
+ \sum_{ {\bf{k}} }\int_{\tau}^{\infty} dt \mbox{        }
\sqrt{ {\bar{n}}_{e}({\bf{k}}){\bar{n}}_{h}({\bf{k}})  }
e^{i(\omega-\epsilon_{ {\bf{k}} })t}
{\tilde{a}}^{\dagger}_{ {\bf{k}} \uparrow }(0,\downarrow,+)
\end{equation}

{\center{ The Optical Conductivity }}

\small
\begin{equation}
 Re(\sigma(\omega)) = |e|\mbox{         } Lt_{ A_{\tau} \rightarrow 0 }
\frac{ cos((\omega-\omega_{X}) \tau) {\tilde{P}}_{I}(\omega,A_{\tau})
 - sin((\omega - \omega_{X})\tau) {\tilde{P}}_{R}(\omega,A_{\tau})
 - cos( (\omega -\omega_{X})\tau) {\tilde{P}}_{I}(\omega,0)
 + sin( (\omega - \omega_{X})\tau) {\tilde{P}}_{R}(\omega,0) }
 { \omega A_{\tau} }
 \end{equation}
 A quick check of dimensions. In units of $ \hbar = c = 1 $, 
 all masses are of dimension inverse length we take to be centeimeter.
 All times are in centimeters. Charge is dimensionless. 
 Since $ \frac{ |e| }{\mu c} A_{\tau} p_{vc}\mbox{   }\tau $ is dimensionless,
 it follows that $ [A_{\tau}] = L^{-1} $. It is easy to check that
 $ [Re(\sigma(\omega))] = L^{-1} $ as it should be.

\normalsize

\section{Results and Discussion}

 The equations written down in the previous sections have to be solved
 numerically. It is worthwhile to point out some pitfalls
 and problems. Let us first focus on the SBEs. 
 As we pointed out in our earlier work\cite{Cartherm}
 involving the SBE's the sums over
 $ {\bf{k}}^{'} $ have to be carried out in a special manner so as
 to avoid potential divergences.
\[
\sum_{{\bf{k \neq k^{'}}}} \frac{ v_{ {\bf{k-k^{'}}} } }{V}
f(k^{'}) =
\frac{(4\pi e^{2})}{(2\pi)^{2}}\int_{0}^{k_{max}} dk^{'}\mbox{ }
(\frac{k^{'}}{k})
ln (\left |\frac{(k^{'}+k)}{(k^{'}-k)} \right |)
(f(k^{'}) - f(k))
\]
\begin{equation}
+ \frac{(4\pi e^{2})}{(2\pi)^{2}}f(k)(\frac{1}{k})
[\frac{1}{2}(k_{max}^{2}-k^{2})
ln (\left |\frac{(k_{max}+k)}{(k_{max}-k)} \right |) + k \mbox{ } k_{max}]
\end{equation}
 Further, it was suggested by Binder et.al. \cite{Binder} that
 we should use a momentum cutoff $ k_{max} = 12/a_{Bohr} $, where
 $ a_{Bohr} $ is the exciton Bohr radius. The justification for this
 stems from the fact that beyond this cutoff the probability of the electron
 existing is negligible. This assertion is true only if the pump field
 frequency is below or equal the band-gap. When the frequency is well
 above the band gap the situation is less clear and care must be taken
 inorder not to lose features that may be present at high momenta. 
 Unfortunately we have found that even when the pump field has a frequency
 equal to the band gap the prescription of Binder et.al\cite{Binder}
 has some problems. In particular, we have found that if we try and sneak a
 peek at the form of the distribution for $ k >> 12/a_{Bohr} $,
 we find a periodic pattern suggesting therefore that electrons
 can exist at (arbitrarily) high momenta long after a pump field
 whose frequency is at the band gap is switched off
 (after 
 a time long enough so that we may still meaningfully talk of a
 well-defined frequency). This is a paradoxical and counterintutive result
 that has been gloosed over by the pioneers \cite{Binder}. The arbitrary
 cutoff of Binder et.al. shold not be taken too seriously. 
 In order to make more sense out of all this we have to claim that 
 the SBE's produce the correct momentum distributions 
 only for small enough $ k $, and we have to use some judgement as to
 where we should cutoff the distributions. 

 The sea-boson analogs of the SBE's written down above have their own
 numerical problems. First is the fact that even in the two-component case
 the sea-boson technique works well only when $ q << k $. 
 Since we have chosen to study only $ {\bf{q}} = 0 $, parts of the
 hamiltonian, it seems that we are in good shape. However we find that
 even then there is a cutoff small $ k = k_{min} $ cutoff below which
 the momentum distributions become unphysical (larger than unity).
 This is true if we use the formula in Eq.(~\ref{NOEQN})
 $ {\bar{n}}_{e}({\bf{k}}) = {\bar{n}}_{h}({\bf{k}}) = {\bar{n}}_{0}({\bf{k}}) $.
 Further we find that this identification is the analog of the SBE.
 It is comforting to know that the sea-boson technique is equivalent
 to the SBE in some limit. In the SBEs, the
 momentum distributions of the electron
 and holes are identical even if the effective masses are very different. This
 is due to the fact that the SBEs neglect the collision terms responsible for
 the asymmetry that we would otherwise expect. Similarly, the
 sea-boson equations at the level of Eq.(~\ref{NOEQN}) neglect
 the repulsion and phonon terms. However the SBEs do include repulsion at 
 the Hartree-Fock level, therefore the analogy between the two is not
 exact. In Fig.1 we see how far we may take this analogy between the SBEs
 and the sea-boson equations. The approach toward 
 unphysical behaviour for the small $ k $ limit of the momentum distribution
 obtained using the sea-boson equations is also seen. When we include the
 effects of repulsion and phonons, the answers change quite dramatically.
 In fact they are so very different from the SBE results that we have decided 
 not to publish them. It will take some more time before a thorough analysis
 is completed and all the ramifications are explored. For now we shall
 assume that the momentum distribution is that given by
 $ {\bar{n}}_{0}({\bf{k}}) $ or that given by the SBEs.
 Let us first write down some formulas that relate the real part of the
 conductivity to the absorption coefficient. We may expect the two to
 have qualitatively similar features. However just to be sure and
 so that we don't make any mistakes having gotten this far, let us write 
 down the formulas. They are a combination of the formulas from 
 the text by Haug and Koch\cite{Haug} and the one by Manah\cite{Mahan}.
 The transverse dielectric function may be decomposed as follows.
\begin{equation}
\epsilon(\omega) = \epsilon_{1}(\omega) + i \mbox{    }\epsilon_{2}(\omega) 
\end{equation}
 The abosrption coefficient, refractive index and the real part of the
 conductivity are given by(in units $ \hbar = c = 1 $),
\begin{equation}
 \alpha(\omega) = \frac{\omega}{n(\omega)}\epsilon_{2}(\omega)
\end{equation}
\begin{equation}
 n(\omega) = \{ \frac{1}{2}
(\epsilon_{1}(\omega)
 + \sqrt{ \epsilon^{2}_{1}(\omega) + \epsilon^{2}_{2}(\omega) }) \}^{\frac{1}{2}}
\end{equation}
The real part of the conductivity is,
\begin{equation}
Re(\sigma(\omega)) = \frac{\omega}{4\pi}
\epsilon_{2}(\omega)
\end{equation}
 It is better not to use the Kramers-Kronig relations as 
 our answer for the real part of the conductivity is undetermined upto
 a factor(actually it has no reason to, it just so happens that the
 magnitude does not agree with observations). We may write down a formula
 for the imaginary part of the conductivity just as we did the real part.
 \small
\begin{equation}
Im(\sigma(\omega)) = |e| Lt_{ A_{\tau} \rightarrow 0 }
\frac{ -sin((\omega-\omega_{X}) \tau) {\tilde{P}}_{I}(\omega,A_{\tau})
 - cos((\omega - \omega_{X})\tau) {\tilde{P}}_{R}(\omega,A_{\tau})
 + sin( (\omega -\omega_{X})\tau) {\tilde{P}}_{I}(\omega,0)
 + cos( (\omega - \omega_{X})\tau) {\tilde{P}}_{R}(\omega,0) }
 { \omega A_{\tau} }
\end{equation}

\normalsize

\begin{equation}
Im(\sigma(\omega)) = -\frac{\omega}{4\pi}
(\epsilon_{1}(\omega)-1)
\end{equation}
 Therefore we may deduce the optical dielectric function and hence the
 absorption coefficient. All this would not be neceesary if the
 refractive index was close to unity and then $ \epsilon_{2} $
 is negligible in comparison with $ \epsilon_{1} $ for all
 frequencies. Then the real part of the conductivity would be 
 proportional to the absorption coefficient.  Let us compare 
 the experimental magnitude of the absorption coefficient and the 
 energy $ \omega $. We find according to the experiments of
 Song's group\cite{Song},  $ |\alpha| \approx 10^{5} cm^{-1} $
 whereas $ \omega \approx 2.0 \pi /(352 \times 10^{-7} cm) = 1.78  \times 10^{5} cm^{-1}  $.
 we can see that these two quantites are comparable to each other
 suggesting thereby that the real part of 
 the dielectric function is not close to unity.

\end{document}